\documentclass{aa}

\usepackage{graphicx}
\usepackage{txfonts}
\usepackage{bm}

\usepackage{natbib}
\bibpunct{(}{)}{;}{a}{}{,}

\newcommand{\beq}{\begin{equation}}
\newcommand{\eeq}{\end{equation}}
\newcommand{\bea}{\begin{eqnarray}}
\newcommand{\eea}{\end{eqnarray}}
\newcommand{\req}[1]{Eq.~(\ref{#1})}
\newcommand{\dd}{\mathrm{d}} 

\newcommand{\EC}{E_\mathrm{C}}
\newcommand{\Ece}{E_\mathrm{ce}}
\newcommand{\Ecp}{E_\mathrm{cp}}
\newcommand{\Eci}{E_\mathrm{ci}}
\newcommand{\Epe}{E_\mathrm{pe}}
\newcommand{\eO}{{\bm{e}}_\mathrm{O}}
\newcommand{\etrv}{\bm{e}^\mathrm{(t)}}
\newcommand{\etr}{{e}^\mathrm{(t)}}
\newcommand{\eX}{{\bm{e}}_\mathrm{X}}

\newcommand{\gcc}{\mbox{g cm$^{-3}$}}

\newcommand{\kB}{k_\mathrm{B}}

\newcommand{\mel}{m_\mathrm{e}} 

\newcommand{\Teff}{T_\mathrm{eff}}
\newcommand{\tEC}{\tilde{E}_\mathrm{C}}
\newcommand{\tEpe}{\tilde{E}_\mathrm{pe}}

\begin{document}

\title{Radiative properties of 
magnetic neutron stars\\
 with metallic surfaces and thin atmospheres}
\titlerunning{Spectra of neutron stars with metallic
surfaces}
                                                         
\author{A. Y. Potekhin
  \inst{1,2,3}
  \thanks{\email{palex@astro.ioffe.ru}}
\and
  V. F. Suleimanov
  \inst{4,5}
\and
  M. van Adelsberg
  \inst{6}
\and
  K. Werner
  \inst{4}
  }

\institute{Centre de Recherche Astrophysique de Lyon
(CNRS, UMR 5574),
Universit\'e Lyon 1,
Ecole Normale Sup\'{e}rieure de Lyon,
46 all\'ee d'Italie,
69364 Lyon Cedex 07, France
\and
Ioffe Physical-Technical Institute,
Politekhnicheskaya 26, St.~Petersburg 194021, Russia
\and
Isaac Newton Institute of Chile, 
         St.~Petersburg Branch, Russia
\and
   Institut f\"ur Astronomie und Astrophysik, Kepler Center for Astro and
Particle Physics, Universit\"at T\"ubingen, Sand 1, 72076 T\"ubingen, Germany
\and
Kazan Federal University, Kremlevskaja Str., 18, Kazan 420008, Russia
\and
   Center for Relativistic Astrophysics, School of Physics, 
   Georgia Institute of Technology, Atlanta, Georgia 30332, USA
}

\date{Received ... / Accepted ...}

\abstract
{Simple models fail to describe
the observed spectra of X-ray-dim isolated neutron
stars (XDINSs). Interpretating these spectra requires detailed
studies of radiative properties
in the outermost layers of neutron stars
with strong magnetic fields. Previous studies
have shown that the strongly magnetized plasma in the outer
envelopes of a neutron star may exhibit a phase transition
to a condensed form.
In this case thermal radiation can emerge directly from the
metallic surface without going through a gaseous
atmosphere, or alternatively, 
it may pass through a ``thin'' atmosphere
above the surface.
The multitude of theoretical possibilities
complicates modeling the spectra and
makes it desirable to have analytic formulae for 
constructing samples of models without going through 
computationally expensive, detailed calculations.
}{
The goal of this work is to 
develop a simple analytic description of 
the emission properties (spectrum and polarization)
of the condensed, strongly magnetized surface of neutron stars.
}{
We have improved the method of 
our earlier work
for calculating the spectral properties of condensed magnetized
surfaces. Using the improved method, we calculated the reflectivity
of an iron surface at magnetic field strengths
$B\sim10^{12}$~G -- $10^{14}$~G, with various inclinations of
the magnetic field lines and radiation beam with respect to
the surface and each other.
We constructed analytic expressions
for the emissivity
of this surface 
as functions of the photon energy,
magnetic field strength,
and the three angles that determine the geometry of the local
problem.
Using these expressions, we calculated X-ray
spectra for neutron stars with condensed iron surfaces covered
by thin partially ionized hydrogen atmospheres.
}{
We develop simple analytic descriptions of the intensity and polarization
of radiation emitted or reflected by 
condensed iron surfaces of neutron stars with the strong magnetic
fields typical of isolated neutron stars. This
description provides boundary conditions at the bottom of a
thin atmosphere, which are more accurate than previously used
approximations. The spectra calculated with this improvement
show different absorption features from those in simplified
models.
}{
The approach developed in this paper yields results that
can facilitate modeling and
interpretation of the X-ray spectra of isolated, strongly
magnetized, thermally emitting neutron stars.
}


\maketitle

\section{Introduction} 
\label{sect:intro}

Recent observations of neutron stars have provided a wealth
of valuable information, but they have also raised many new
questions. Particularly intriguing is the class of
radio-quiet neutron stars with thermal-like spectra,
commonly known as X-ray dim isolated neutron stars (XDINSs),
or the Magnificent Seven (see, e.g., reviews by
\citealt{Haberl07} and \citealt{Turolla09}, and references
therein). Some of them (e.g., RX J1856.5$-$3754) have
featureless spectra, whereas others (e.g., RX J1308.6+2127
and RX J0720.4$-$3125) have broad absorption features with
energies $\sim0.2$\,--\,2 keV. In recent years,
an accumulation of observational evidence has suggested 
that XDINSs may
have magnetic fields $B\sim10^{13}$\,--\,$10^{14}$~G and be
related to magnetars (e.g., \citealt{Mereghetti08}). 

For interpretating the XDINS spectra, it may be necessary to
take the phenomenon of ``magnetic condensation'' into
account.
The strong magnetic field squeezes the electron clouds around the
nuclei, thereby increasing the binding and cohesive energies
(e.g., \citealp{MedinLai06}, and references therein). Therefore
XDINSs may be ``naked,'' with no appreciable atmosphere above a
condensed surface, as first conjectured by
\citet{ZaneTD02}, or they may have  a
relatively thin atmosphere, with the spectrum of outgoing
radiation affected by the properties of
the condensed surface beneath the atmosphere, 
as suggested by \citet{MotchZH03}.

Reflectivities of condensed metallic surfaces in strong
magnetic fields have been studied in several papers 
\citep{Brinkmann80,TurollaZD04,vanAdelsberg-ea05,PerezAMP05}. 
\citet{Brinkmann80} and \citet{TurollaZD04} neglected  the
motion of ions in the condensed matter, whereas 
\citet{vanAdelsberg-ea05} (hereafter Paper~I)  and
\citet{PerezAMP05} considered two opposite limiting cases,
one that neglects the ion motion (``fixed ions'') and
another where the ion response to the electromagnetic wave
is treated by neglecting the Coulomb interactions between the
ions (``free ions''). A large difference between these two
limits occurs at photon frequencies below the ion cyclotron
frequency, but the two models lead to almost the same
results at higher photon energies. We expect that in reality
the surface spectrum lies between these two limits (see
Paper~I for discussion). The results of Paper~I  and of
\citet{PerezAMP05} are similar, but
differ significantly from the earlier results. In
particular, \citet{TurollaZD04} find that collisional
damping in the condensed matter leads to a sharp cutoff in
the emission at low photon energies, but such a cutoff is
absent in Paper~I and \citet{PerezAMP05}. It is most
likely that this difference arises from the ``one-mode''
description for the transmitted radiation adopted by 
\citet{TurollaZD04} (see Paper~I for details). All the
previous works relied on a complicated method of finding the
transmitted radiation modes, originally due to
\citet{Brinkmann80}. We replace it with the more reliable method
described below. 

\citet{Ho-ea07} (see also \citealp{Ho07})
fitted multiwavelength 
observations of RX J1856.5$-$3754 
with a model of a thin, magnetic, partially ionized hydrogen
atmosphere on top of a condensed iron surface; they also
discuss possible mechanisms of creation of such a thin
atmosphere. \citet{SuleimanovPW09} calculated
various models of fully and partially ionized
finite atmospheres above a condensed surface
including the case of ``sandwich'' atmospheres,
composed of hydrogen and helium layers above a condensed surface.

The wide variety of theoretical possibilities complicates the
modeling and interpretation of the spectra. To
facilitate this task, \citet{Suleimanov-ea10} 
           (hereafter Paper~II)
suggest an approximate treatment, in which the local
spectra, together with temperature and magnetic field
distributions, are fitted by simple analytic functions. By being
flexible and fast, this approach is suited to constrain stellar
parameters prior to performing more accurate, but computationally
expensive calculations of model spectra. The
reflectivity of the condensed surface was modeled by a simple
steplike function, which roughly described
the polarization-averaged reflectivity of a
magnetized iron surface at $B=10^{13}$~G,
but depended neither on the magnetic field
strength $B$ nor on the angle $\varphi$ between the plane of
incidence and the plane made by the normal to the surface and the
magnetic field lines.

In the present work,  the numerical method of Paper~I and
the approximate treatment of Paper~II are refined.  We
develop a less complicated and more stable method of
calculations and construct more accurate fitting formulae
for the reflectivities of a condensed, strongly magnetized
iron surface, taking the dependence on arguments $B$ and
$\varphi$ into account. The new fit reproduces the feature
near the electron plasma energy, obtained numerically in
Paper~I but neglected in Paper~II. Two versions of the fit
are presented in Sect.~\ref{sect:refl} for the models of
free and fixed ions discussed in Paper~I. In addition to the
fit for the average reflectivity, we present analytic
approximations for each of the two polarization modes, which
allow us to calculate the polarization of radiation of a
naked neutron star. In Sect.~\ref{sect:atm} we consider the
radiative transfer problem in a finite atmosphere above the
condensed surface, including the reflection from the inner
atmosphere boundary with normal-mode transformations,
neglected in the previous studies of thin atmospheres.
Conclusions are given in Sect.~\ref{sect:concl}. In
Appendix~\ref{sect:Improved} we describe the method of
calculation for the reflectivity coefficients, which is
improved with respect to Paper~I. In Appendix~\ref{sect:rmj}
we describe an analytic model of normal-mode reflectivities
at the inner boundary of a thin atmosphere.

\section{Spectral properties of
a strongly magnetized neutron star surface} 
\label{sect:refl}

\subsection{Condensed magnetized surface}
\label{sect:surface}

Most of the known neutron stars have much larger magnetic fields $B$
than the natural atomic unit
for the field strength $B_0=e^3 \mel^2 c/\hbar^3 = 
2.35\,\times 10^9$~G, which is set by equating the electron cyclotron
energy
\beq
   \Ece = \hbar eB/\mel c =115.77~B_{13}\textrm{ keV}
\label{omce}
\eeq
to the Hartree unit of energy $\mel e^4/\hbar^2$.
Here, $\mel$ is the electron mass, $e$ the elementary
charge, $c$ the speed of light in vacuum, 
$\hbar$ the Planck constant divided by $2\pi$,
and $B_{13}=B/10^{13}$~G. 
Fields with $B \gg B_0$ profoundly affect the properties
of atoms, molecules, and plasma (see, e.g., \citealp{NSB1},
chap.~4). 
\citet{Ruderman71} suggested that the strong magnetic field
may 
stabilize linear molecular chains (polymers) aligned with the
magnetic field and eventually
turn the surface of a neutron star into the metallic
solid state.
Later studies have provided support for this conjecture,
although 
the surface density $\rho_\mathrm{s}$ and, especially,
the critical temperature $T_\mathrm{crit}$
below which such condensation
occurs remain uncertain.
Order-of-magnitude estimates suggest
\beq
   \rho_\mathrm{s} = 
   8.9\times10^3\,\eta\,AZ^{-0.6}\,B_{13}^{1.2}
   \textrm{~~~\gcc},
\label{rho_s}
\eeq
where $A$ and $Z$ are the atomic mass and charge numbers,
and $\eta\sim1$ an unknown numerical factor, which
absorbs the theoretical uncertainty \citep[see][]{Lai01}. The
value $\eta=1$ corresponds to the equation of state
provided by the ion-sphere model \citep{Salpeter54}.
More recent results of the zero-temperature Thomas-Fermi
model for $^{56}$Fe at
$10^{10}\mbox{~G}\leqslant B\leqslant 10^{13}$~G
\citep{Fushiki-ea,Rognvaldsson-ea} can be  approximated 
(within 4\%) by \req{rho_s} with
$\eta\approx0.2+0.0028/B_{13}^{0.56}$, whereas the
finite-temperature Thomas-Fermi model of
\citet{Thorolfsson-ea} does not predict magnetic
condensation at all. The most comprehensive study of
cohesive properties of the magnetic condensed surface has
been conducted by \citet{MedinLai06,MedinLai07}, based on
density-functional theory (DFT).
\citet{MedinLai06} calculated cohesive energies
$Q_\mathrm{s}$ of the molecular chains and condensed
phases of H, He, C, and Fe in strong magnetic fields. A
comparison with previous DFT calculations by
other authors suggests that $Q_\mathrm{s}$ may vary within a
factor of two at $B\gtrsim10^{12}$~G, depending on the
approximations employed (see \citealp{MedinLai06} for references and
discussion). \citet{MedinLai07} calculated equilibrium
densities of saturated vapors of He, C, and Fe atoms and
chains above the condensed surfaces and  obtained
$T_\mathrm{crit}$ at several values of $B$ by equating the
vapor density to $\rho_\mathrm{s}$. Unlike previous
authors, \citet{MedinLai06,MedinLai07} have taken the
electronic band structure of the metallic phase into account
self-consistently. However, in the gaseous phase, they
still did not allow for atomic motion across the magnetic
field and did not take a detailed treatment 
of excited atomic
and molecular states into account. \citet{MedinLai07} calculated the
surface density assuming that the linear molecular chains
(directed along $\bm{B}$) form a rectangular array in the
perpendicular plane and that the distance between the nuclei
along the field lines is the same in the condensed matter
as in the separate molecular chain
\citep{MedinLai06,Medin12}. \citet{MedinLai07} found that
the critical temperature is $T_\mathrm{crit}\approx0.08
Q_\mathrm{s}/\kB$. Their numerical results for  $^{56}$Fe at
$0.5\leqslant B_{13} \leqslant 100$ can be described by expression 
$T_\mathrm{crit} \approx (5+2\,B_{13})\times10^5$~K for the
critical temperature and by \req{rho_s} with $\eta\approx0.55$ for
the surface density, with uncertainties below 20\% for both
quantities. 
  An observational determination of the phase state of a
neutron star surface would be helpful for improving the
theory of matter  in strong magnetic fields.

The density of saturated vapor above the condensed surface
rapidly decreases with decreasing $T$ \citep{LaiSalpeter97}.
Therefore, although the surface is hidden by an optically
thick atmosphere at $T\approx T_\mathrm{crit}$, the
atmosphere becomes optically thin at $T\ll T_\mathrm{crit}$.
Also, as  suggested by \citet{MotchZH03}, there may be a
finite amount of light chemical elements (e.g., H) on top of
the condensed surface of a heavier element (e.g., Fe). In
addition,  the same atmosphere may be optically thick for
low photon energies and transparent at high energies. The
energy at which the total optical thickness of a finite
atmosphere equals unity depends on the atmosphere column
density, which can in turn depend on temperature. At a fixed
energy, the optical thickness of the finite atmosphere is
different for different photon polarizations, therefore the
atmosphere can be thick for one polarization mode and thin
for another. One should take all these possibilities into
account while interpreting observed spectra of neutron
stars.

\subsection{Formation of the spectrum}
\label{sect:Formation}

\subsubsection{Normal modes and polarization vectors}
\label{sect:NM}

It is well known \citep[e.g.,][]{Ginzburg} that under
typical conditions (e.g., far from the resonances) 
electromagnetic radiation propagates in a magnetized plasma in
the form of extraordinary (X) and ordinary (O) normal
modes.  These modes have different polarization vectors
$\eX$ and $\eO$, absorption
and scattering coefficients, and refraction and
reflection coefficients at the surface.
\citet{GP73} studied conditions for the
applicability of the normal-mode description
and formulated the radiative transfer
problem in terms of these modes.

Following the works of \citet{Shafranov} and
\citet{Ginzburg}, \citet{HoLai01} derived convenient
expressions for the normal mode polarization vectors in a
fully ionized plasma for photon energies $E$ much higher
than the electron plasma energy
\beq
   \Epe =\left({4\pi\hbar^2 e^2 n_\mathrm{e} 
                      / \mel} \right)^{1/2}
                      \approx 0.0288\sqrt{\rho\, Z/A} 
                      \textrm{~~keV},
\label{Epe}
\eeq
where $\rho$ is the density in \gcc.
In the complex representation of plane waves with 
$\bm{E} \propto \bm{e}\,
\mathrm{e}^{\mathrm{i}(\bm{k}\cdot\bm{r}-\omega t)}$,
in the coordinate system where the $z$-axis is along the
wave vector $\bm{k}$, and the magnetic field $\bm{B}$ lies
in the $(xz)$ plane, 
the polarization vectors
are
\beq
   \bm{e}_M(\alpha) = \frac{1}{\sqrt{1+|K_M(\alpha)|^2
 + |K_{z,M}(\alpha)|^2}}
 \left( \begin{array}{c}
     \mathrm{i} K_M(\alpha) \\  1  \\  \mathrm{i} K_{z,M}(\alpha)
     \end{array}
     \right),
\label{e'}
\eeq
We use the notation $M=\mathrm{X}$ and $M=\mathrm{O}$ for
the extraordinary and ordinary polarization modes,
respectively. $K_M(\alpha)$ and $K_{z,M}(\alpha)$ are
functions of the angle $\alpha$ between $\bm{B}$ and
$\bm{k}$. They are determined by the dielectric tensor of
the plasma and thus depend on the photon energy $E$, as
well as $\rho$, $B$, $T$ and the chemical composition.
\citet{HoLai03} calculated $K_M$ and studied the polarization
of normal modes including the effect
of the electron-positron vacuum polarization, while
\citet{Potekhin-ea04} additionally considered an incomplete
ionization of the plasma.

\begin{figure}
\begin{center}
\includegraphics[width=.6\columnwidth]{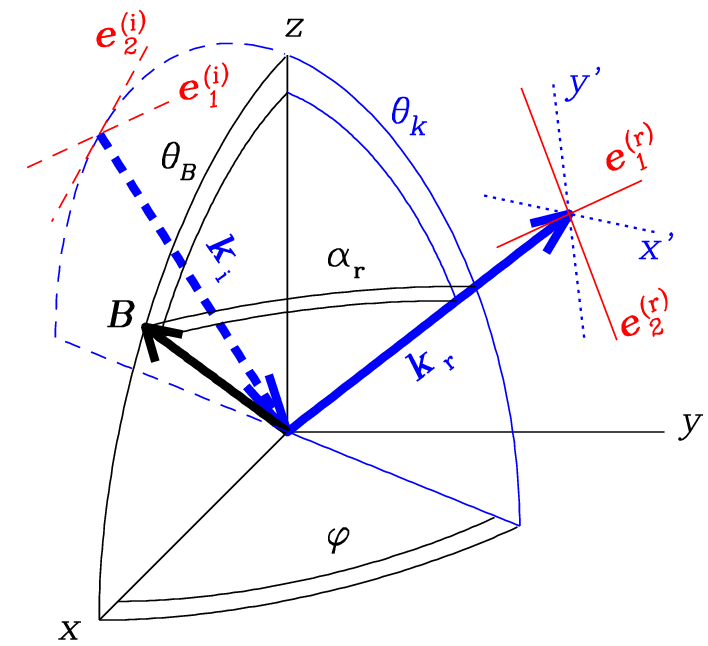}
\end{center}
\caption{Illustration of notations. The $z$ axis is chosen
perpendicular to the surface, and the $(xz)$ plane is chosen parallel
to the magnetic field lines,
which make an angle $\theta_B$ with the normal to the surface.
The direction of the reflected beam with wave vector
$\bm{k}_\mathrm{r}$
is determined by the polar
angle $\theta_k$ and the azimuthal angle $\varphi$, and
$\alpha_\mathrm{r}$ is
the angle between the reflected beam and the field lines.
Thick solid lines show the reflected beam and magnetic field
directions, thin solid lines illustrate the coordinates,
and dashed lines show the incident photon wave vector
$\bm{k}_\mathrm{i}$ and  its
quadrant. The lines marked $\bm{e}_{1,2}^\mathrm{(i,r)}$ illustrate
the basic polarizations adopted for the description 
of reflectivities: $\bm{e}_{1,2}^\mathrm{(i)}$ and $\bm{e}_{1,2}^\mathrm{(r)}$ are perpendicular to
 the wave vectors $\bm{k}_\mathrm{i}$ and $\bm{k}_\mathrm{i}$, respectively;
 $\bm{e}_1^\mathrm{(i,r)}$ are parallel to the surface, and
 $\bm{e}_2^\mathrm{(i,r)}$ lie in the perpendicular plane.
 The axes $x'$ and $y'$ lie in the plane made by 
 $\bm{e}_1^\mathrm{(r)}$ and $\bm{e}_2^\mathrm{(r)}$, 
$x'$ being aligned in the plane made by
$\bm{B}$ and $\bm{k}_\mathrm{r}$.
}
\label{fig:plot1}
\end{figure}

\subsubsection{Emission and reflection by a condensed surface}

The condition $E > \Epe$ is usually satisfied for X-rays
in neutron star atmospheres, but not in the
condensed matter. We consider a 
surface element that is sufficiently small for the variation
in the magnetic field strength and inclination to be neglected.
We treat this small patch as plane, neglecting
its curvature and roughness. We choose
the Cartesian $z$ axis perpendicular to this plane
and the $x$ axis parallel to the projection of
magnetic field lines onto the $xy$ plane. We denote
the angle between the field and the $z$ axis as $\theta_B$,
the incidence angle of the radiation as $\theta_k$, and the 
angle in the $xy$ plane made by the projection of the 
wave vector as $\varphi$ (Fig.~\ref{fig:plot1}). 
The
angle between the wave vector and magnetic field lines is
given by
\beq
   \cos\alpha_\mathrm{i,r} =
     \sin\theta_B \sin\theta_k  \cos\varphi
      \mp 
     \cos\theta_B \cos\theta_k
\label{alpha}
\eeq
for the incident and reflected waves, respectively.
The surface emits radiation 
with monochromatic intensities
\beq
I_{E, j}= J_j\,B_E / 2
\quad
(j=1,2).
\eeq
Here, the basis for polarization is chosen such that 
the waves with $j=1$ and 2 are linearly polarized parallel 
and perpendicular to the incident plane, respectively 
(Fig.~\ref{fig:plot1});
$I_{E, j}\,\dd\Sigma\,\dd\Omega\,\dd E\, \dd t$ gives the
energy radiated in the $j$th wave by the surface element
$\dd\Sigma$, in the energy band $(E,E+\dd E)$, during time
$\dd t$, in the solid angle element $\dd\Omega$ around the
direction of the wave vector $\bm{k}$.  We use the function  
\beq
   B_E = \frac{B_\nu}{2\pi\hbar}=
   \frac{E^3}{4\pi^3\hbar^3c^2(\mathrm{e}^{E/\kB T}-1)},
\eeq
where $B_\nu$ is Planck's spectral radiance and $\kB$
the Boltzmann constant. Dimensionless emissivities for
the two polarizations, normalized to blackbody values,
are $J_j = 1-R_j$, where $R_j$ is the effective
reflectivity of mode $j$ defined in Appendix~\ref{sect:Improved}. 
The reflectivities depend on surface material, photon energy, 
magnetic field strength $B$ and inclination $\theta_B$, and on 
the direction of $\bm{k}$.  For nonpolarized radiation, it is
sufficient to consider the mean reflectivity
$
   R = (R_1+R_2)/{2}
$
(see Paper~I).

\subsubsection{Reflectivity calculation}
\label{sect:reflGen}

A method for calculating reflectivity coefficients was
developed in Paper~I. However, it is not easy to implement.
Though it mostly produces correct results, in some ranges of model
parameters it can yield unphysical results, which are
difficult to distinguish from the correct ones. In the
present work we present an improved method that
avoids this complication (see Appendix~\ref{sect:Improved}).

Using our new method, we calculated the spectral properties
of a condensed Fe surface and compared the results with those in Paper~I. 
As in Paper~I, we considered two alternative models for the 
response of ions to electromagnetic waves in the condensed phase:
one neglects the Coulomb interactions
between ions, while the other treats ions as
frozen at their equilibrium positions in the Coulomb crystal
(i.e., neglecting their response to the electromagnetic wave).

\begin{figure}
\begin{center}
\includegraphics[width=\columnwidth]{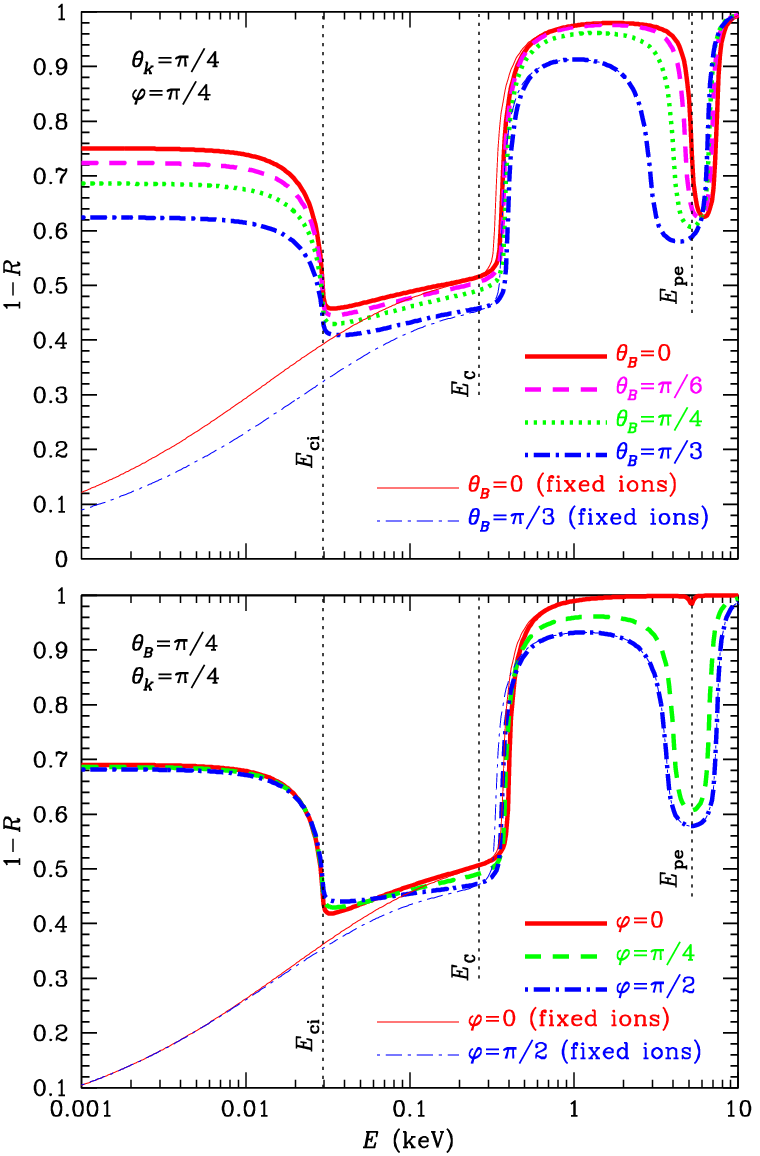}
\end{center}
\caption{Dimensionless emissivity $J=1-R$  as a function of
photon energy $E$ for a condensed Fe surface with  $B =
10^{13}$~G and $T=10^6$~K.
The top panel shows several cases with varying
$\theta_B$ and fixed $\theta_k = \pi/4$,  $\varphi = \pi/4$.
The bottom panel shows several cases with varying $\varphi$
and fixed $\theta_k = \pi/4$, $\theta_B = \pi/4$.  These plots
should be compared with Figs.~5 and 6 of 
Paper~I.}
\label{fig:Correct}
\end{figure}

In the first limiting case (thick lines in
Fig.~\ref{fig:Correct}), the reflectivity exhibits
different behavior in three characteristic energy ranges: $E <
\Eci$, $\Eci \leqslant E \lesssim \EC$,  and $E \gtrsim \EC$, where
\beq
      \Eci = {\hbar ZeB}/{A m_\mathrm{u} c}
   =0.0635\,({Z}/{A})\, B_{13}
   \textrm{~keV}
\eeq
is the ion cyclotron energy, $m_\mathrm{u}$ 
is the unified atomic mass unit, and
\beq
  \EC=\Eci+\Epe^2/\Ece.
\eeq
In addition, there is suppression of the reflectivity at
$E\sim\Epe$; the exact position, width, and depth of the suppression 
depend on the geometry defined by the angles $\theta_B$, $\theta_k$, and
$\varphi$.

In the opposite case of immobile ions (thin curves in
Fig.~\ref{fig:Correct}), the reflectivity has a similar
behavior at $E>\Eci$, but differs at $E \lesssim \Eci$. It
does not exhibit the sharp change at $E\approx\Eci$, but
smoothly continues to the lower energies. As argued in
Paper~I, we expect that the actual reflectivity lies
between these two extremes.

The new results, shown in Fig.~\ref{fig:Correct}, display
the same qualitative behavior as in Paper~I,
but exhibit considerable deviation from the previous
calculations for some geometric settings in the energy range
$\Eci \lesssim E \lesssim \EC$. Thus, the qualitative
results and conclusions of Paper~I are correct, but the new 
method described in Appendix~\ref{sect:Improved} is
quantitatively more reliable.

If $\theta_B=\theta_k=0$, then an approximate analytic solution 
(neglecting the finite electron relaxation rate in the medium; see
Paper~I for discussion) is $R\approx (R_+ + R_-)/2$, where
\beq
   R_\pm^{(0)} = \left|\frac{n_\pm^{(0)}-1}{n_\pm^{(0)}+1}\right|^2,
\quad
   n_\pm^{(0)} = \left[
    1 \pm \frac{\Epe^2}{\Ece(E\pm\Eci)}
     \right]^{1/2}.
\label{Rj0}
\eeq
Compared to the numerical results, \req{Rj0} provides a good
approximation at $E\lesssim\Eci$. Therefore, we use it in the
analytic fit described below.

\subsection{Results for iron surface}
\label{sect:reflFe}

In the numerical examples presented below we 
assume a condensed $^{56}$Fe surface and use the
estimate of the surface density given by the ion-sphere model --
that is, we set $\eta=1$ in \req{rho_s}.

\subsubsection{Mean reflectivity}

\begin{figure}
\includegraphics[width=\columnwidth]{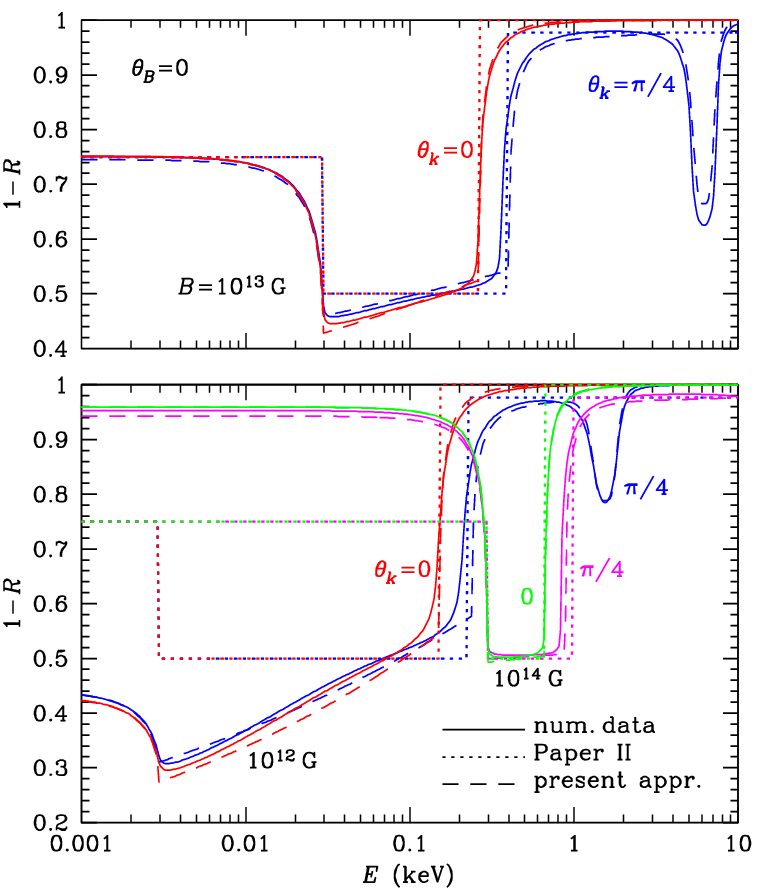}
\caption{
Dimensionless emissivity $J=1-R$ as a function of photon energy
$E$ for condensed Fe surface at $B=10^{13}$~G (top panel),
$B=10^{12}$~G and $10^{14}$~G (bottom panel), with magnetic field
lines normal to the surface, for two angles of incidence
$\theta_k=0$ and $\theta_k=\pi/4$, as marked near the curves.
Solid lines show our numerical results, and dashed lines
demonstrate the fit. For comparison, dotted lines reproduce the
simplified approximation used in Paper~II.
}
\label{fig:rt0}
\end{figure}

In practice, the average normalized emissivity $J=1-R$ is
usually more important than the specific emissivities $R_j$.
In Paper~II, $R(E)$ was replaced  by a constant in each of
the three ranges mentioned in Sect.~\ref{sect:reflGen}
($E<\Eci$, $\Eci<E<\tEC\approx\EC$, and $E>\tEC$).
For simplicity, the values of these three constants were
assumed to depend only on $\theta_B$ and $\theta_k$, but not
on $\varphi$ or $B$. Here we propose a more elaborate and
accurate fit, which is a function of $E$,
$B$, $\theta_B$, $\theta_k$, and $\varphi$, for the
magnetic field range $10^{12}\mbox{~G}\lesssim B \lesssim
10^{14}$~G and photon energy range $\mbox{1~eV}\lesssim E
\lesssim 10$~keV.  In the approximation of free ions, the
average reflectivity of the metallic iron surface is
approximately reproduced by
\beq
   J = \left\{ \begin{array}{l}
   J_\mathrm{A}
    \quad
    \mbox{in Region~I},
    \\
    \displaystyle
    J_\mathrm{B}\,(1-J_\mathrm{C})    
     + \frac{J_\mathrm{C}}{1+{L}}
    \quad\textrm{in Region~II}.
    \end{array} \right.
\label{fit}
\eeq
Region~I is the low-energy region defined by the conditions
$E<\Eci$ and $J_\mathrm{A} > J_\mathrm{B}$. 
Region~II is the supplemental range of relatively high
energies in which either of these conditions is violated.
The functions $J_\mathrm{A}$, $J_\mathrm{B}$, and $J_\mathrm{C}$
are mainly responsible for the behavior 
of the emissivity at $E < \Eci$, $\Eci < E \lesssim \tEC$,
and $E > \tEC$, respectively, while the function ${L}$
describes the line at $E\approx\Epe$.  The value 
\beq
   \tEC = \Eci + \tEpe^2/\Ece
\label{EC1}
\eeq
is the energy at which
the square of the effective refraction index 
\beq
  \tilde{n}^2 =  
    1 - \frac{\tEpe^2}{\Ece(E-\Eci)}
\label{n12a}
\eeq
(analogous to \req{Rj0})
becomes positive with increasing $E$
in the range $E>\Eci$.
In Eqs.~(\ref{EC1}) and (\ref{n12a}),
\beq
   \tEpe = \Epe \sqrt{
      3-2\cos\theta_k
        }.
\eeq

The low-energy part of the fit in \req{fit} is given by
\beq
   J_\mathrm{A}=[1-A(E)]\,J_0(E),
\eeq
where
\beq
  A(E) = \frac{1-|\cos\theta_B|}{2\sqrt{1+B_{13}}}\,
       +  \left[
       0.7-\frac{0.45}{J_0(0)}\,
       \right]
          \,(\sin\theta_k)^4\,(1-\cos\alpha),
\label{A}
\eeq
$J_0(E)=1-\frac12(R_-^{(0)}+R_+^{(0)})$,
and $R_\pm^{(0)}$ are given by \req{Rj0}.
Accordingly, 
$J_0(0) = 4\left(\sqrt{\EC/\Eci}+1\right)^{-1}
\left(\sqrt{\Eci/\EC}+1\right)^{-1}$.
In \req{A} and hereafter, $\alpha$ without subscripts denotes
$
   \min(\alpha_\mathrm{r},\alpha_\mathrm{i}).
$

In the intermediate energy range, $\Eci<E\lesssim\tEC$,
there is a wide suppression of the emissivity. We describe this
part by the power-law interpolation between the values at
$\Eci$ and $\tEC$:
\beq
   J_\mathrm{B} = (E/\tEC)^p J(\tEC),
\quad\textrm{where~~}
   p=\frac{\ln[J(\tEC)/J(\Eci)]}{\ln(\tEC/\Eci)}.
\label{JB}
\eeq
The values $J(\Eci)$ and $J(\tEC)$ are approximated as follows:
\bea
   J(\tEC) &=& \frac12
    + \frac{0.05}{1+B_{13}}
      (1+|\cos\theta_B|\sin\theta_k)
\nonumber\\&&
        - 0.15(1-|\cos\theta_B|)\sin\alpha,
\\
   J(\Eci) & = & \frac{2 n_0}{(1+n_0)^2}\,
         \left(
      1 + \frac{|\cos\theta_B|-\cos\theta_k}{2\,(1 + B_{13})}
         \right),
\eea
where
$
   n_0 = ( 1 + {\Epe^2}/{2\Ece\Eci} )^{1/2}.
$

The steep slope at $E>\tEC$ is described by \req{Rj0}
with $\Epe$ replaced by $\tEpe$:
\beq
   J_\mathrm{C} = \left\{ \begin{array}{l}
       4\tilde{n}/(1+\tilde{n})^2
         \quad\mbox{ at } E > \tEC, \\
         0\quad\mbox{ at } E \leqslant \tEC,
      \end{array} \right.
\label{JC}
\eeq
$\tilde{n}$ being given by \req{n12a}.

Finally, the lowering
of $J(E)$ at $E>\tEC$ is fit by
\bea
   {L} &=& \left[
    \frac{0.17\Epe/\EC
          }{1+X^4}
        +0.21\,\mathrm{e}^{-(E/\Epe)^2}
        \right] (\sin\theta_k)^2\,W_L,
\label{L}
\\&&
   X = \frac{E-E_L}{2\Epe W_L}\,(1-\cos\theta_k)^{-1},
\nonumber\\&&
   E_L=\Epe\left[ 1+1.2\,(1-\cos\theta_k)^{3/2} \right]
      \,\left[ 1-(\sin\theta_B)^2/3 \right],
\nonumber\\&&
   W_L = 0.8\,(\tEC/\Epe)^{0.2}
     \sqrt{\sin(\alpha/2)}\, \left[ 1+(\sin\theta_B)^2 \right].
\nonumber\eea
The line at $E_L$ disappears from the fit 
($L\to0$) when radiation is parallel to the
magnetic field ($\alpha\to0$). This property is not exact;
our numerical results reveal a remnant of the 
line at $\alpha\to0$, which is relatively weak,
but may become appreciable
if the magnetic field is strongly inclined ($\theta_B>\pi/4$).

\begin{figure}
\includegraphics[width=\columnwidth]{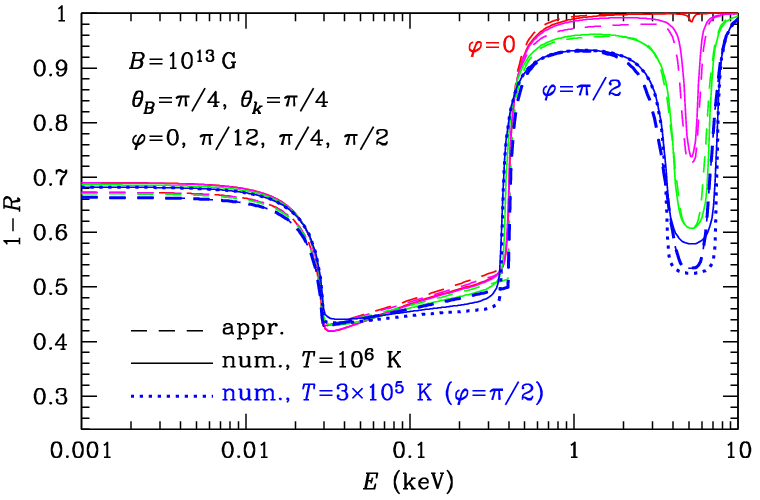}
\caption{
Dimensionless emissivity $J=1-R$ as a function of photon energy
$E$ for a condensed Fe surface with inclined magnetic field
($B=10^{13}$~G, $\theta_B=\pi/4$) and
inclined incidence of radiation ($\theta_k=\pi/4$ and four values
of $\varphi$ listed in the figure). 
Solid lines show our numerical results for the model of free ions
at $T=10^6$~K, 
and short-dashed lines
demonstrate the fit.
For comparison, the dotted line reproduces
our numerical results for $\varphi=\pi/2$ and $T=3\times10^5$~K.
}
\label{fig:rb1t2i2}
\end{figure}
\begin{figure}
\includegraphics[width=\columnwidth]{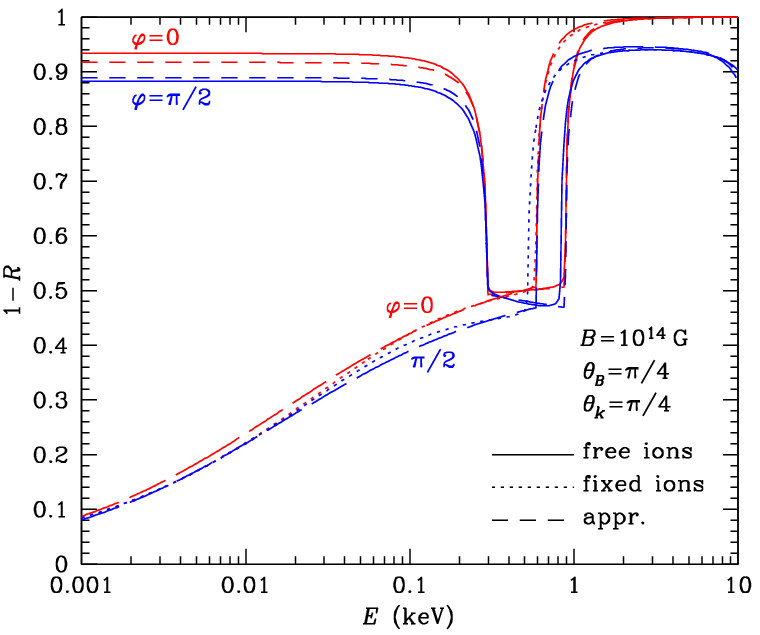}
\caption{
Same as in Fig.~\ref{fig:rb1t2i2}, but for $B=10^{14}$~G
and two values of $\varphi$. For comparison,
the dotted and long-dashed lines reproduce
our numerical results and analytic approximation,
respectively,
for the model of fixed ions.
}
\label{fig:rb4t2i2}
\end{figure}
\begin{figure}
\includegraphics[width=\columnwidth]{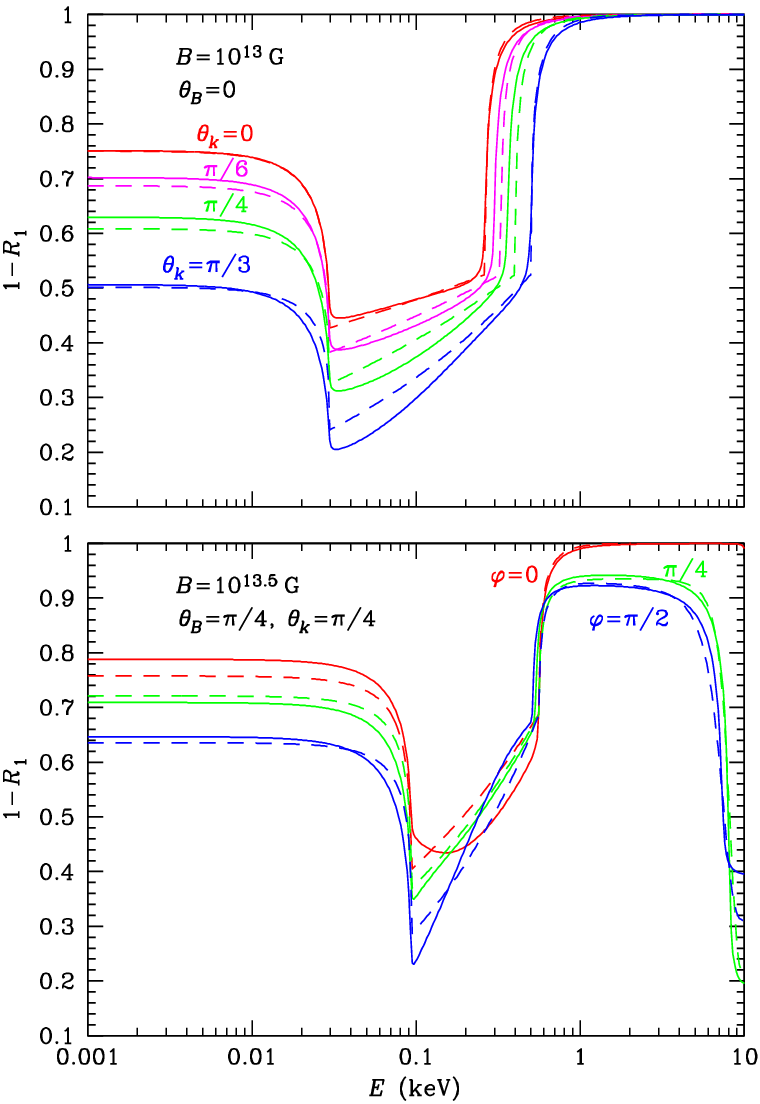}
\caption{
Dimensionless emissivity for the linear polarization $\bm{e}_1$,
$J_1=1-R_1$, as a function
of photon energy $E$ for condensed Fe surface at different
magnetic field strengths and geometric settings. 
Top panel:
magnetic field $B=10^{13}$~G is normal to the surface,  and
angles of radiation incidence are $\theta_k=0$, $\pi/6$, $\pi/4$, and
$\pi/3$. 
Bottom panel: 
 $B=10^{13.5}$~G, magnetic field lines and the
photon beam are both inclined at $\theta_B=\theta_k=\pi/4$,
and the azimuthal angle takes values $\varphi=0$, $\pi/4$, and
$\pi/2$.
Solid lines show the
numerical results, and dashed lines
demonstrate the fit.
}
\label{fig:rmod1}
\end{figure}
\begin{figure}
\includegraphics[width=\columnwidth]{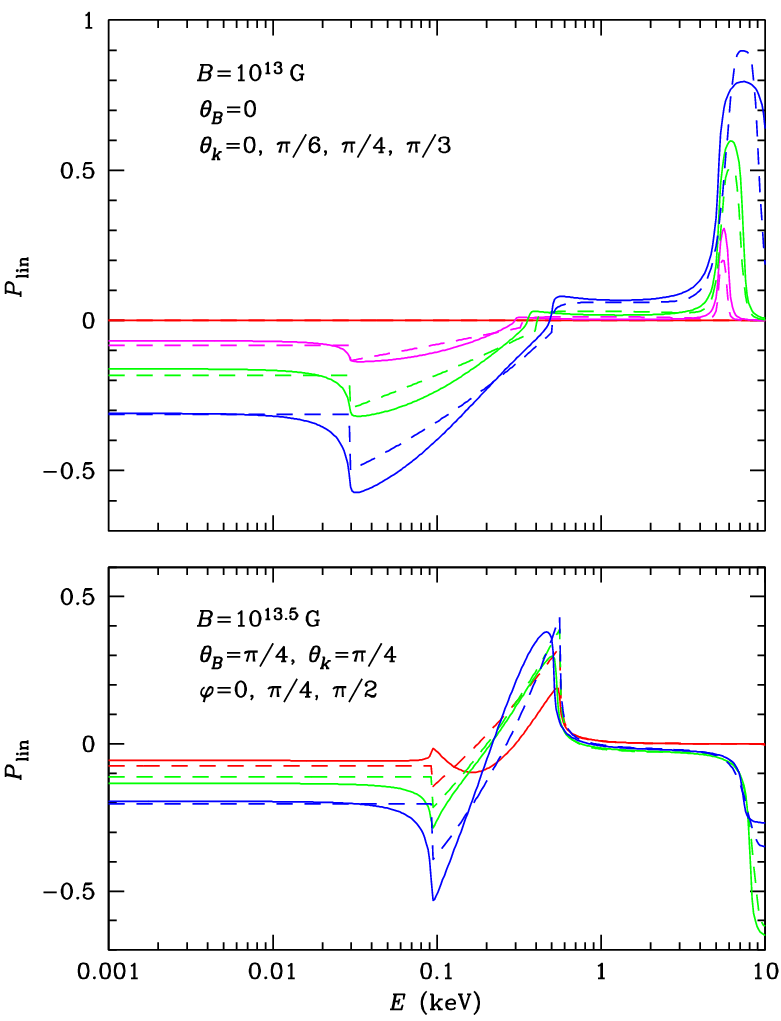}
\caption{
Degree of linear polarization $P_\mathrm{lin}$ [\req{P}] as a
function of photon energy $E$ for condensed Fe surface.
The values of $B$, directions of the field and the wave vector,
and line types
are same as in Fig.~\ref{fig:rmod1}.
}
\label{fig:polar}
\end{figure}

Examples of the numerical results for the normalized 
emissivities are compared to the analytic approximation in
Figs.~\ref{fig:rt0}\,--\,\ref{fig:rb4t2i2}. For most
geometric settings, the fit error lies within 10\% in more
than 95\% of the interval $-3 < \log_{10}E\textrm{ (keV)} <
1$. The remaining $<5$\% are the narrow ranges of $E$ where
$J(E)$ sharply changes. Exceptions occur for
strongly inclined fields ($\theta_B>\pi/4$) and small $\varphi$,
where the error may exceed 20\% in up to
10\% of the logarithmic energy range.

Our fit does not take the dependence of
reflectivity on temperature into account. Temperature of the condensed
matter enters in the calculations through the effective
relaxation time, which determines the damping factor (see
Paper~I). This disregard is justified by the weakness of the
$T$-dependence of the results.  With decreasing $T$, the
transitions of $R(E)$ between characteristic energy ranges
become sharper, and the feature at $E_L$ becomes stronger.
The bulk of our calculations employed in the fitting was
done at $T\sim10^6$~K. In Fig.~\ref{fig:rb1t2i2}, which
shows the numerical results at $T=10^6$~K, 
an additional line is drawn at $T=3\times10^5$~K, 
in order to illustrate the $T$-dependence.

A small modification of the proposed approximation can describe
the alternative model of fixed ions (see Paper~I). In this
case, it is sufficient to formally set $\Eci\to0$ in the
above equations and replace \req{JB} by
\beq
   J_\mathrm{B} = \frac{J(\tEC)
      }{
         1-p+p\,(\tEC/E)^{0.6}
         },
\quad\textrm{where~~}
   p=0.1\,\frac{1+\sin\theta_B}{1+B_{13}}.
\label{JBfix}
\eeq
As an example, in Fig.~\ref{fig:rb4t2i2} the
numerical results
in the model of fixed ions and the
fit \req{JBfix} are shown in addition to the free-ion
results.

\subsubsection{Polarization}

We also constructed approximations for the
emissivities in each of the two modes. Their functional
dependence on $E$ and geometric angles in Fig.~\ref{fig:plot1}
is more complicated than the analogous dependence for the average
$J=(J_1+J_2)/2$. We did not accurately reproduce these
complications, in order to keep the fit relatively simple,
but reproduced general trends. For $j=1$, our free-ion
approximation has the same form as \req{fit}:
\beq
   J_1 = \left\{ \begin{array}{l}
   J_\mathrm{A1}
    \quad\textrm{in Region~I},
    \\
    J_\mathrm{B1}\,(1-J_\mathrm{C})    
     + J_\mathrm{C} (1 - R_L)
    \quad\textrm{in Region~II}.
    \end{array} \right.
\label{fit1}
\eeq
Here, we retain $J_\mathrm{C}$ given by \req{JC}. 
The shape of the line near the plasma
frequency is unchanged and is described by \req{L},
but the line strength
is different, since $L$ enters
\req{fit1} through the function
\bea
      R_L &=& 
      (\sin\theta_B)^{1/4}\,\left[2-(\sin\alpha)^4 \right]
      \,\frac{{L}}{1+{L}} .
\eea
The  functions that describe emissivity in mode 1 at $E <
\tEC$ are
\bea
   J_\mathrm{A1} &=& [ 1 - A_1 ] \, J_\mathrm{A},
\\&&
   A_1 = \frac{a_1}{1+0.6\,B_{13}\,(\cos\theta_B)^2},
\nonumber\\&&
   a_1 = 1-(\cos\theta_B)^2\cos\theta_k-(\sin\theta_B)^2\cos\alpha;
\nonumber\\
   J_\mathrm{B1} &=& (E/\tEC)^{p_1} J_1(\tEC),
\qquad
   p_1=\frac{\ln[J_1(\tEC)/J_1(\Eci)]}{\ln(\tEC/\Eci)},
\label{JB1}\\&&
   J_1(\Eci) = (1-a_1)\,J(\Eci),
\nonumber\\&&
   J_1(\tEC) = \frac12 +
   \frac{0.05}{1+B_{13}}+\frac{\sin\theta_B}{4}.
\nonumber
\eea
In the fixed-ion case, it is sufficient to set $\Eci\to0$
and to replace \req{JB1} by
\beq
    J_\mathrm{B1} = \frac{J_1(\tEC)
       }{
          0.1+0.9\,(\tEC/E)^{0.4}} .
\eeq
For the second mode, no additional fitting is needed,
because 
$R_2=2R-R_1$ and $J_2=2J-J_1$.

Figure~\ref{fig:rmod1} compares the use of
Eqs.~(\ref{fit1})\,--\,(\ref{JB1})
to numerical results. The upper
panel shows the case where the field lines are perpendicular to
the surface. In this case the line at $E_L$ disappears from 
mode 1, so the line in $R$ seen in Fig.~\ref{fig:rt0} for
$\theta_k\neq0$ is entirely due to mode 2. As soon as the field
is inclined, the line is redistributed between the two modes (the
lower panel of Fig.~\ref{fig:rmod1}). In the latter case the
numerical results show a more complex functional
dependence $R_1(E)$ in the range $\Eci<E<\tEC$, which is not
fully reproduced by our fit, for the reasons 
discussed above.

The azimuthal angle $\varphi$ enters the fit only
through $\alpha$. As a consequence, the fit is symmetric with
respect to a change in sign of $\varphi$. This property
may seem natural at first glance; however, we note that
the numerical results do not strictly obey this symmetry, which
holds for the nonpolarized beam, but not for each of the
polarization modes separately. We have checked that this is not a numerical
artifact: because the magnetic field vector $\bm{B}$ is
axial, there is no strict symmetry with respect to the
$(x,z)$ plane. A reflection about this plane would require 
simultaneous inversion of the $\bm{B}$ direction in order
to restore the original results. However, as long as the
electromagnetic waves are nearly transverse (i.e., $K_z$ in
(\ref{e'}) and (\ref{et}) are small), the asymmetry is weak,
allowing us to ignore it and thus keep the fit
relatively simple.

The analytic approximations in Eqs.~(\ref{fit}) and (\ref{fit1}) 
allow one to evaluate the degree of linear polarization of the emitted
radiation 
\beq
   P_\mathrm{lin} = (J_1-J_2)/2J = (R_2-R_1)/(2-2R).
\label{P}
\eeq
For example, the two panels of Fig.~\ref{fig:polar} show
$P_\mathrm{lin}$ for 
the same directions of the magnetic field and the photon beam
as in the respective panels of Fig.~\ref{fig:rmod1}.
We see that the analytic formulae, originally devised to
reproduce the normalized emissivities, also reproduce 
the basic features of $P_\mathrm{lin}(E)$. Although the feature
at $E\sim\Epe$ is absent in the top panel of
Fig.~\ref{fig:rmod1}, it reappears in the top panel of
Fig.~\ref{fig:polar} due to the contribution of $R_2$ in \req{P}.

\section{X-ray spectra of thin atmospheres}
\label{sect:atm}

\subsection{Inner boundary conditions}
\label{sect:inner}

Propagation of radiation in an atmosphere is described by
two normal modes (see Sect.\,\ref{sect:NM}).  At the inner
boundary of a thin atmosphere, an incident X-mode beam of
intensity $I_{E}^{\mathrm{X}}$ gives rise to reflected beams
in both modes, whose intensities are proportional to
$I_{E}^{\mathrm{X}}$, and analogously for an incident
O-mode. Therefore,  the inner boundary conditions for
radiation transfer in an atmosphere of a finite thickness
above the condensed surface can be written as
\bea
   I_{E}^{\mathrm{X}}(\theta_k,\varphi) &=& 
   \textstyle\frac12
      J_\mathrm{X}(\theta_k,\varphi)
      B_E(T)
       + R_\mathrm{XX}(\theta_k,\varphi)\,
        I_{E}^{\mathrm{X}}(\pi-\theta_k,\varphi)
\nonumber\\&&
        + R_\mathrm{XO}(\theta_k,\varphi)\,
        I_{E}^{\mathrm{O}}(\pi-\theta_k,\varphi),
\label{XX}
\\
   I_{E}^{\mathrm{O}}(\theta_k,\varphi) &=& 
   \textstyle\frac12
      J_\mathrm{O}(\theta_k,\varphi)
      B_E(T)
       + R_\mathrm{OO}(\theta_k,\varphi)\,
        I_{E}^{\mathrm{O}}(\pi-\theta_k,\varphi)
\nonumber\\&&
        + R_\mathrm{OX}(\theta_k,\varphi)\,
        I_{E}^{\mathrm{X}}(\pi-\theta_k,\varphi),
\label{OO}
\eea
where $I_{E}^M$  ($M=\mathrm{X,O}$) 
are the specific intensities of 
the X- and O-modes in the atmosphere 
at $\rho=\rho_\mathrm{s}$,
$R_{MM'}$ are 
coefficients of reflection with allowance for 
transformation of the incident mode $M'$ into the reflected mode
$M$,
and $J_M$ are the normalized emissivities.
The latter can be written by analogy with $J_{1,2}$ as
$J_\mathrm{X}=1-R_\mathrm{X}$ and $J_\mathrm{O}=1-R_\mathrm{O}$,
where $R_\mathrm{X}=R_\mathrm{XX}+R_\mathrm{XO}$
and $R_\mathrm{O}=R_\mathrm{OO}+R_\mathrm{OX}$
(cf.\ Paper~I). 

\citet{Ho-ea07} retained only the emission terms $\frac12 J_M B_E$ on
the right-hand sides of Eqs.\,(\ref{XX}), (\ref{OO}). The
reflection was taken into account in Paper~II, but
calculations were performed neglecting $R_\mathrm{OO}$,
$R_\mathrm{OX}$, and $R_\mathrm{XO}$, under the
assumption that $R_\mathrm{XX}$ is equal to $R$ and does
not depend on $\varphi$. Here
we use a more realistic,
albeit still approximate, model for $R_{MM'}$, described in
Appendix~\ref{sect:rmj}.

\subsection{Results}
\label{sect:res}

Here, we illustrate the importance of the correct description
of the reflection for computations of thin model
atmospheres above a condensed surface. To this end, we have
calculated a few model atmospheres with normal magnetic
field (therefore, $\theta_B  =\varphi = 0$, and 
$\alpha_\mathrm{r} = \alpha_\mathrm{i} = \theta_k$),  taking
the model with $B = 4\times 10^{13}$\, G, effective
temperature $\Teff$ = 1.2 $\times 10^6$\,K,  and surface
density $\Sigma = 10$ g cm$^{-2}$ as a fiducial model. In
the fiducial model the free-ions assumption for  
condensed-surface reflectivity is used.

\begin{figure}
\includegraphics[width=\columnwidth]{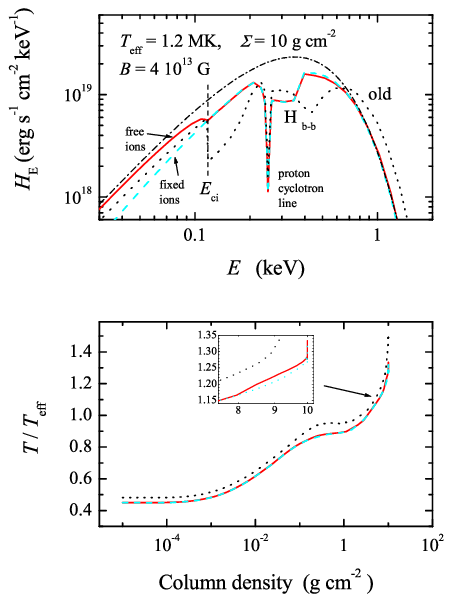}
\caption{ Emergent spectra (top panel) and temperature structures (bottom panel) for the fiducial model 
atmosphere (solid curve) and for model atmospheres that are
calculated using the fixed-ions approximation for
 the reflectivity calculations (dashed curves), and  the 
 inner boundary condition from Paper II (dotted curves). In
the top panel the diluted blackbody spectrum that
 fits the high-energy part of the fiducial model spectrum  is also shown (dash-dotted curve).
 }
\label{fig:sp1}
\end{figure}
\begin{figure}
\includegraphics[width=\columnwidth]{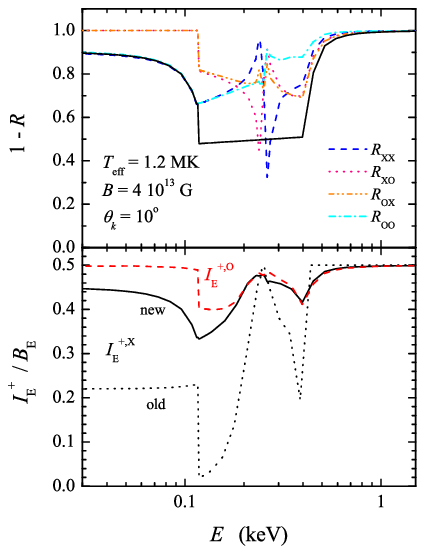}
\caption{Top panel: dimensionless emissivities for
coefficients of reflection $R_\mathrm{XX}$ (dashed curve),
$R_\mathrm{XO}$ (dotted curve), $R_\mathrm{OX}$
(dash-dot-dotted curve), and $R_\mathrm{OO}$ (dash-dotted
curve). The quantities are calculated at the bottom of the
fiducial model atmosphere for the angle between the radiation
propagation and magnetic field, $\theta_k$ = 10$\degr$, 
together with the total  dimensionless emissivity (solid
curve). Bottom panel: dimensionless outward specific
intensities (inner boundary condition) at the  bottom of the
fiducial model atmosphere for the X-mode (solid curve) and
O-mode (dashed curve).  For comparison, the dotted curve
shows  the same for the X-mode, calculated using the  inner
boundary condition from Paper II (in this case  the
dimensionless specific intensity of the O-mode equals 0.5).
}
\label{fig:refl}
\end{figure}
\begin{figure}
\includegraphics[width=\columnwidth]{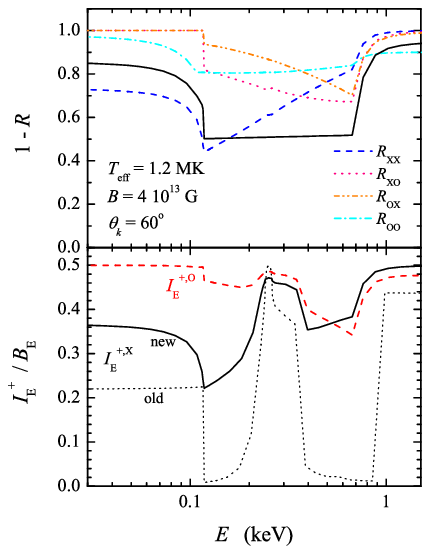}
\caption{The same as in Fig.\,\ref{fig:refl},
but for $\theta_k$ = 60$\degr$.
}
\label{fig:refl2}
\end{figure}

For these computations we use the numerical code
described in \cite{SuleimanovPW09}, with a modified
iterative procedure for temperature corrections. We evaluate
these corrections using
the Uns\"old-Lucy method (e.g., \citealt{Mihalas}),
which gives a better
convergence for thin-atmosphere models than 
other standard methods. In our case, the deepest atmosphere  point
is the upper point of the condensed surface. The
temperature
correction at this point is
obtained as follows: the total flux at the boundary between the
atmosphere and the condensed surface is fixed and,
therefore, the following energy balance condition has to be
satisfied:
\bea \label{ec}
H_0 &=& \frac{\sigma_\mathrm{SB}\Teff^4}{4\pi}
 = \frac{1}{2}\int_0^\infty\,\dd E\, 
\int_{-1}^{1} \left(
 I_{E}^{\mathrm{X}}(\mu) +I_{E}^{\mathrm{O}}(\mu)
   \right)\,\mu\, \dd\mu 
\nonumber \\
&=& B_\mathrm{tot}\,k_\mathrm{RL} + JR + H^-.
\eea
Here, $\sigma_\mathrm{SB}$ is the
Stefan-Boltzmann constant, $\mu=\cos \theta_k$, and
\bea \label{ec1}
B_\mathrm{tot} &=& \int_0^\infty B_E\,\dd E\,, 
\nonumber \\
k_\mathrm{RL} &=& \frac{1}{2\,B_\mathrm{tot}}
   \int_0^\infty B_E\,\dd E
     \int_0^1 (1-R)\,\mu\,\dd\mu, 
\nonumber \\
JR &=& \frac{1}{2}\,\int_0^\infty\, \dd E\, 
\nonumber \\
&\times&\int_0^1\, \left( I_{E}^{\mathrm{X}}(\mu)(R_\mathrm{XX}+R_\mathrm{OX}) +
I_{E}^{\mathrm{O}}(\mu)(R_\mathrm{XO}+R_\mathrm{OO})\right)
\,\mu\,\dd\mu, 
\nonumber \\
H^- &=& \frac{1}{2}\int_0^\infty\,\dd E \, 
\int_{-1}^{0} \left(
 I_{E}^{\mathrm{X}}(\mu) +I_{E}^{\mathrm{O}}(\mu)
   \right)\,\mu\, \dd\mu.
\eea

Generally, the condition (\ref{ec}) is not fulfilled at a
given temperature iteration. Therefore,
we perform a linear expansion of the integrated blackbody intensity:
\beq \label{db}
H_0 =(B_\mathrm{tot}+\Delta B_\mathrm{tot})k_\mathrm{RL} + JR + H^-,
\eeq
 and find a corresponding temperature correction
\beq
\label{dt}
\Delta T =
\frac{\pi}{4\sigma_\mathrm{SB}T^3}\left(\frac{1}{k_\mathrm{RL}}\, 
\left(H_0 -
B_\mathrm{tot}\,k_\mathrm{RL}-JR-H^-\right)\right).
\eeq  
This last-point correction procedure is stable and has a convergence rate similar to 
the Uns\"old-Lucy procedure at other depths.

We also changed the depth grid for a better description
of the temperature structure in thin-atmosphere models. 
In semi-infinite model atmospheres that do not have a 
condensed surface as a lower boundary, a logarithmically
equidistant set of depths is used. However, in thin-atmosphere 
models, such a set yields 
insufficient accuracy at the
boundary between the atmosphere and condensed surface.
To improve the description of the boundary, we divide 
the model atmosphere into two parts with equal thicknesses
and use logarithmically equidistant depth grids for each of
them. In the upper part the grid starts from outside (the
closest points are at the smallest depths), 
while in
the lower part it starts from the condensed surface (the 
closest points are at the deepest depths).
This combined grid allows us to describe the whole
atmosphere with the desired accuracy of 1\% for the
integral flux conservation.

In Fig.\,\ref{fig:sp1} we show the emergent spectra and temperature structures for
three different model atmospheres with the same
fiducial set of physical parameters.  The model spectrum computed
using the inner boundary 
condition described in Paper II (the ``old model'') significantly differs
from the two other model spectra computed using the
improved boundary conditions of Eqs.~(\ref{XX}), (\ref{OO}).
The latter two models are calculated using the
free (fiducial model) and fixed ions (alternative model) 
assumptions for the condensed surface reflectivities.

The differences between the spectra and the temperature
structures of these two models are very small. The
atmosphere temperature near the condensed surface with fixed
ions is slightly smaller than the temperature near the
condensed surface with free ions. The flux in the spectrum
of the fiducial model is  approximately twice that of the
alternative model at photon energies $E$ smaller than the iron
cyclotron energy $\Eci =0.118$~keV. At larger energies
the spectra are very close to each other. We note that the
old model has been computed using the free-ions
assumption.   

\begin{figure}
\includegraphics[width=\columnwidth]{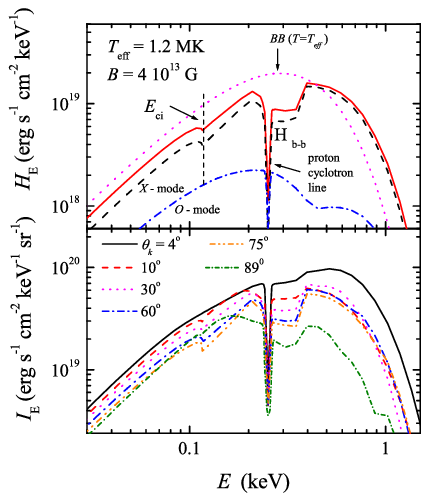}
\caption{ {Top panel:}
total emergent spectrum of the fiducial model (solid curve),
together with emergent spectra in the X-mode (dashed curve)
and in the O-mode (dash-dotted curve). The blackbody
spectrum with $T=T_{\rm eff}$ is also shown (dotted curve).
Bottom panel: emergent specific intensities  of the fiducial
model for six angles $\theta$. 
}
\label{fig:sp2}
\end{figure}
\begin{figure}
\includegraphics[width=\columnwidth]{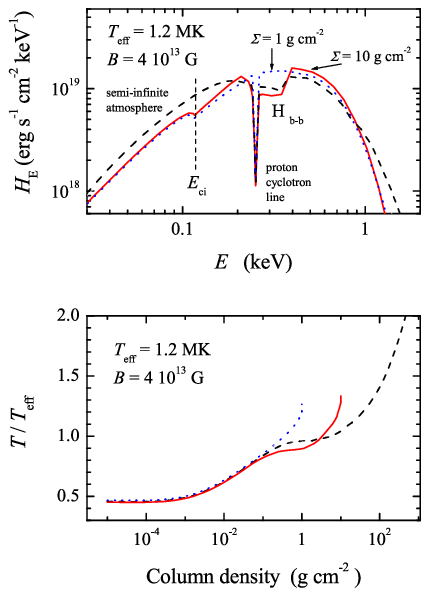}
\caption{Comparison of emergent spectra (top panel) and temperature structures (bottom panel) of the fiducial
model (solid curves) with the semi-infinite model atmosphere (dashed curves) and with the thinner ($\Sigma$ 
= 1 g cm$^{-2}$) model atmosphere (dotted curves), but with the same magnetic field. 
}
\label{fig:sp3}
\end{figure}
\begin{figure}
\includegraphics[width=\columnwidth]{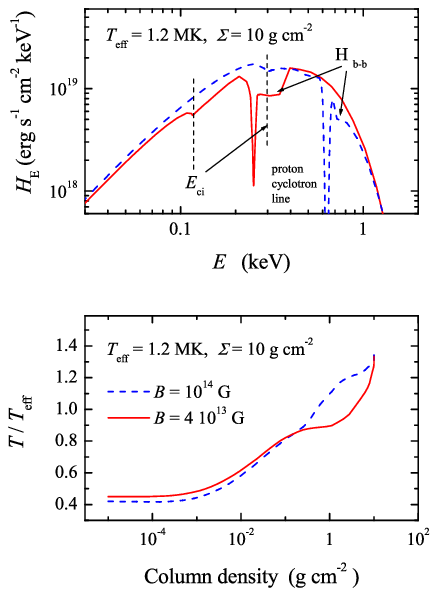}
\caption{Comparison of emergent spectra (top panel) and
temperature structures (bottom panel) of the fiducial model
(solid curves) and  the model atmosphere with different
magnetic field ($B$ = 10$^{14}$ G), but with  the same
surface density (dashed curves).
}
\label{fig:sp4}
\end{figure}

The difference in the emergent spectra between the old and
new model atmospheres is significant. In the old model,
there is a deep depression of the spectrum between $\Eci$
and $\EC$ with an emission-like feature around the
absorption line at the proton cyclotron energy $\Ecp =
0.252$~keV. In the new spectra this complex feature
between   $\Eci$  and $\EC$ is completely different. The
total depression is not significant, but instead of the flux
increase, there appears a deep absorption feature at photon
$E\gtrsim\Ecp$. This  absorption corresponds to the
bound-bound transitions in hydrogen atoms in strong magnetic
field  (H$_\mathrm{b-b}$ feature). 

It is clear that this difference arises due to the inner
boundary condition. The bottom panels of 
Figs.\,\ref{fig:refl} and \ref{fig:refl2} illustrate the
difference in the outgoing flux at the boundary between the
atmosphere and the condensed surface for the old and new
models. This difference is especially large for the flux in
the X-mode. In the old model we assumed complete
reflection in the X-mode; therefore, the reflected flux was 
small as the atmosphere was optically thin at these
energies. As a result we found a small emergent flux at these
energies. In the new models, we have significant mode
transformation due to reflection, which causes an
appreciable part of the energy from the O-mode to convert
into the X-mode; the converted photons then almost 
freely escape from the atmosphere. The reflectivity 
coefficients $R_{MM'}$ for two angles $\theta_k$ are shown 
in the upper panels of Figs.\,\ref{fig:refl} and \ref{fig:refl2}. 

The total equivalent widths (EWs) of the complex absorption
features in the spectra of the new models are smaller  than
the EW of this feature in the spectrum of the old model.
Nevertheless, they are still significant, with EW $\approx$
220\,--\,250 eV, if the continuum is assumed to be a diluted
blackbody spectrum that fits the high-energy tail of the
model. The parameters of the diluted blackbody  spectrum are
a color correction factor $f_c = T/\Teff = 1.2$, and a
dilution factor $D = 1.1^{-4}$.  The range of values for EW
is sufficient to explain the observed absorption features of
XDINSs (for reviews, see \citealt{Haberl07,Turolla09}).

The new and old spectra are strongly polarized, with most of
the  energy radiated in the X-mode (see Fig.\,\ref{fig:sp2},
upper panel). 
We note that, for the parameters of the fiducial model, the
vacuum resonance density occurs between the X and O mode
photon decoupling densities.  Therefore, the polarization
signal does not exhibit a rotation of the plane of
polarization between low and high energies. In contrast,
models that exhibit this effect, considered by \citet{LaiHo03} and
\citet{vanAdelsbergLai06}, have a
lower magnetic field and higher effective temperature,
causing the vacuum resonance to occur outside the X and O
photospheres.

The angular distribution of the
emergent flux is different in the two models
(Fig.\,\ref{fig:sp2}, bottom panel),  especially at photon
energies between $\EC$ and 4$\EC$. In the old model, the
angular distribution is peaked around the surface normal. In
the new model, the emergent radiation is almost isotropic,
with a peak  around the surface normal at the broad 
H$_\mathrm{b-b}$ absorption feature.

The influence of the atmosphere thickness on its emergent
spectrum and the temperature structure is illustrated in
Fig.\,\ref{fig:sp3}.  A thinner atmosphere with $\Sigma$ = 1
g cm$^{-2}$ has an insignificant  H$_\mathrm{b-b}$
absorption feature because it is formed at higher column
densities ($\approx1$\,--\,2  g cm$^{-2}$). The fiducial
model has the smallest temperature at these column densities
among all the models. As a result, the H$_\mathrm{b-b}$
absorption feature is most significant in the spectrum of
this model. The spectrum of the semi-infinite atmosphere has
a hard tail and does not have any feature at $\Eci$.

The importance of the H$_\mathrm{b-b}$ absorption feature
decreases if it is located far from the maximum of the 
spectrum. This is illustrated by the comparison of the
fiducial model with the model calculated for $B = 10^{14}$\,
G (Fig.\,\ref{fig:sp4}). In the latter case, the
H$_\mathrm{b-b}$ absorption feature is less visible and
cools the atmosphere at column densities about a few g
cm$^{-2}$ less efficiently,  although the EW decreases
insignificantly. 

\subsection{Discussion: toward models of observed spectra}

Our calculations are presented for a local patch of the
neutron-star surface with particular values of $\Teff$ and
$\bm{B}$. By taking surface distributions of $\Teff$ and
$\bm{B}$ into account, one can construct an emission
spectrum from the entire neutron star; however this spectrum
is necessarily model-dependent, as the $\Teff$ and $\bm{B}$
distributions are generally unknown. If these distributions
are sufficiently smooth, then integration over the surface
makes absorption features broader and shallower, as
demonstrated, e.g., in the case of cooling neutron stars
with dipole magnetic fields and semi-infinite
\citep{Ho-ea08} or thin (Paper~II) partially ionized
hydrogen atmospheres. As shown in Paper~II, smearing of the
features is stronger, if the crustal magnetic field has a
strong toroidal component, but weaker, if radiation is
formed at small hotspots on the surface, where $\bm{B}$ can
be considered as constant. Using the results of Paper~II,
\citet{Hambaryan-ea11} fitted observed phase-resolved
spectra of XDINS RBS 1223 and derived constraints on
temperature and magnetic field strength and distribution
in the X-ray emitting areas, their geometry, and the
gravitational redshift at the surface. The present, more
detailed approximations for the reflectivities can be
directly used to refine these fits and constraints.

In our numerical examples presented above, we evaluated the
density of the condensed matter using \req{rho_s} with
$\eta=1$. An eventual correction to this approximation is
rather straightforward, once $\rho_\mathrm{s}$ is accurately
known. Indeed, the density enters calculations through
$\Epe\propto\sqrt{\eta}$ [\req{Epe}] and through the damping
factor (Paper~I). The latter dependence is relatively weak,
therefore, it is sufficient to correct $\Epe$ in the
expressions presented in Sect.~\ref{sect:refl}.

As mentioned in Sect.~\ref{sect:res}, the model spectra are
highly polarized at the stellar surface. However, the
observed polarization signal is affected by propagation of
the photons through the neutron-star magnetosphere. In
addition to redshift and light bending effects near the
stellar surface, the mode eigenvectors evolve adiabatically
along with the direction of the changing magnetic field in
the magnetosphere \citep[see, e.g.,][]{HeylShaviv02}. The
adiabatic evolution continues until the photons near the
polarization limiting radius, $r_{\rm pl}$, which is
typically many stellar radii from the surface.  At $r_{\rm
pl}$, the modes couple, with the intensities and
eigenvectors frozen thereafter (in addition, significant
circular polarization can be generated in some cases, for
example, in radiation from rapidly rotating neutron stars;
see \citealp{vanAdelsbergLai06}, and the references
therein).

The main effect of adiabatic photon propagation in the
magnetosphere is on the synthetic polarization signal from a
finite region of the neutron-star surface. (There is an
additional effect for photons propagating through a
quasi-tangential region of magnetic field near the stellar
surface; in the majority of cases this ``QT effect'' can be
neglected -- see \citealp{WangLai09} for details). Since the
mode properties are fixed at large distances from the star,
where the magnetic field is aligned for photon trajectories
from different areas on the star, the polarization signal is
not as diminished due to variation in the surface magnetic
field as might be expected if vacuum polarization effects
are ignored \citep[see][]{HeylShaviv02,Heyl-ea03}. 
Thus, it is possible that polarization features of the local
thin atmosphere models described above may be retained in
spectra from a finite region of the neutron star.  Observed spectra
and polarization signals have been presented in the
literature, employing several atmosphere models for
emission from the entire surface \citep{Heyl-ea03} and
from a finite sized hotspot \citep{vanAdelsbergPerna09}.

\section{Conclusions} 
\label{sect:concl}

We have improved the method of Paper~I
for calculating spectral properties of condensed magnetized 
surfaces.
Using the improved method, we
calculated a representative set of reflectivities of a metallic
iron surface for the magnetic field strengths
$B=10^{12}$~G\,--\,$10^{14}$~G. Based on these calculations, we
constructed  analytic expressions for emissivities of the
magnetized condensed surface in the two normal modes as functions
of five arguments: energy of the emitted X-ray photon $E$, field
strength $B$, field inclination $\theta_B$, and the two angles
that determine the photon direction. We considered the
alternative limiting approximations of free and fixed ions for 
calculating the condensed surface reflectivity.

We improved the inner boundary conditions for the
radiation transfer equation in a thin atmosphere above a
condensed surface. The new boundary condition accounts for
the transformation of normal modes into each other caused by 
reflection from the condensed surface. To implement this
condition we suggested a method for calculating 
reflectivities $R_{MM'}$ in the normal modes used for model
atmosphere calculations, based on analytic
approximations to the reflectivities.

We computed a few models of thin, partially ionized hydrogen
atmospheres to investigate the influence of the new boundary
condition on their emergent spectra and temperature
structures. The allowance for mode transformations makes the
complex absorption feature between $\Eci$ and $\EC$ less
significant and the atomic absorption feature more
important.  Nevertheless, the equivalent widths of this
complex absorption feature in the emergent spectra are still
significant ($\approx200$\,--\,250 eV) and sufficient to
explain the observed absorption features in the spectra of
XDINSs. Models of thin atmospheres with inclined magnetic
fields are necessary for detailed descriptions of their
spectra. We plan to compute such models with vacuum
polarization and partial mode conversion in a future paper.

\begin{acknowledgements}
A.Y.P.\ acknowledges useful discussions with Gilles
Chabrier, the hospitality of the Institute of Astronomy and
Astrophysics at the University of T\"ubingen, and partial
financial support from Russian Foundation for Basic Research
(RFBR grant no.\,11-02-00253-a) and the Russian Leading
Scientific Schools program (grant NSh-4035.2012.2).  The
work of V.F.S.~is supported by the German Research
Foundation (DFG) grant SFB/Transregio 7 ``Gravitational Wave
Astronomy''. 
\end{acknowledgements}
 

\appendix
\section{Improved reflectivity calculation}
\label{sect:Improved}

In this Appendix, we describe several improvements to the methods
of Paper~I that have enabled us to produce a general, efficient
code, free of numerical difficulties, which computes the correct
value of the reflectivities  over the full range of parameters
used in neutron-star atmosphere modeling.

In general,
each incoming linearly polarized wave
$\bm{E}_1^{\mathrm{(i)}}=\mathcal{A}_1\bm{e}_1^{\mathrm{(i)}}$
 and $\bm{E}_2^{\mathrm{(i)}}=\mathcal{A}_2\bm{e}_2^{\mathrm{(i)}}$
is partially reflected, giving rise to reflected and
transmitted
fields\footnote{We use the inverse
order of the subscripts in $r_{mj}$ and $t_{mj}$
with respect to the one used
in Paper~I.}
\beq
 \bm{E}^\mathrm{(r)}_j = \mathcal{A}_j \sum_{m=1}^2
  r_{mj} \,\bm{e}^\mathrm{(r)}_m,
\qquad
 \bm{E}^\mathrm{(t)}_j = \mathcal{A}_j \sum_{m=1}^2
  t_{mj} \,\bm{e}^\mathrm{(t)}_m .
\label{def_r}
\eeq
As shown in Paper~I, 
the dimensionless emissivities for the two orthogonal linear
polarizations are $J_j=1-R_j$, where
\beq
   R_j = |r_{j1}|^2+|r_{j2}|^2
\quad
   (j=1,2).
\label{Rj-def}
\eeq

The reflected field amplitudes $r_{11}, r_{12}, r_{21}$, and
$r_{22}$ were calculated in Paper~I using an eighth-order
polynomial in the refraction index $n_j$ to determine the  properties of the
transmitted modes. The transmitted wave can be
described by two normal modes, thus, most of the  roots obtained
from that polynomial represent unphysical solutions to the
equations. The conditions to identify the correct
roots were derived in Appendix~B of Paper~I,
with the requirements that the
corresponding reflectivities satisfy $R_1, R_2 \leqslant 1$
and that the function  $R(E)$ be continuous.
However, for some values of the model
parameters, the  unphysical roots can satisfy the
physical constraints on the solution. 

Here we propose an improved method, based on a fourth-order
polynomial, which allows for easy elimination of the unphysical
roots.

\subsection{Transmitted modes}

A significant simplification of the equations describing the
transmitted mode properties can be obtained by writing the 
transmitted wave vector as
\begin{eqnarray}
\label{eq:wavevec}
\bm{n}_j = \frac{c}{\omega}\bm{k}_j & = n_j
            \left[\sin\theta_j\left(\cos\varphi\,{\hat\mathbf{x}}
            +  \sin\varphi\,{\hat\mathbf{y}}\right) +
            \cos\theta_j\,{\hat\mathbf{z}}\right]
\nonumber\\
& =
\sin\theta_k\left(\cos\varphi\,{\hat\mathbf{x}}
 +\sin\varphi\,{\hat\mathbf{y}}\right)
+ n_{z,j}\,{\hat\mathbf{z}},
\end{eqnarray}
where ${\mathbf{\hat{x}}}$, ${\hat\mathbf{y}}$, 
${\hat\mathbf{z}}$ are unit
vectors along the $x$, $y$, $z$ axes, respectively, and the 
quantities $\cos\theta_j$,  $\sin\theta_j$ are complex
numbers satisfying the condition: $\cos^2\theta_j +
\sin^2\theta_j = 1$ (cf.\ Appendix~B of  Paper~I).  The
second equality in \req{eq:wavevec} follows from 
Snell's Law, $n_j\sin\theta_j = \sin\theta_k$, and the 
definition $n_{z,j} \equiv n_j \cos\theta_j$.  

From Maxwell's equations for the transmitted modes,
\begin{eqnarray}
\label{eq:MEs}
{\bm\lambda}\cdot \bm{E} & = & \mathbf{0},\\
{\bm\lambda} & \equiv & {\bm\epsilon} + \bm{n}_j\otimes\bm{n}_j -
n^2_j\,\mathbf{I},
\end{eqnarray}  
where ${\bm\epsilon}$ is the dielectric tensor of the medium
(see Eq.~(13) of Paper~I),  $\mathbf{I}$ is the unit
tensor, and $\bm{E}$ is the electric field vector.  If we
note that  $n^2_j =
n^2_j\left(\sin^2\theta_j+\cos^2\theta_j\right) =
\sin^2\theta_k + n^2_{z,j}$, and apply the condition 
$\det{{\bm\lambda}} = 0$ to obtain a nontrivial solution to
\req{eq:MEs}, the result is a fourth-order polynomial in
$n_{z,j}$:
\begin{eqnarray}
\label{eq:polynj}
a_4 n^4_{z,j} & + & a_3 n^3_{z,j} + a_2 n^2_{z,j} + a_1
n_{z,j} + a_0 = 0,\\
a_4 & = & 1 + \sin^2\theta_k/\left(\epsilon_{zz} -
\sin^2\theta_k\right),\\
a_3 & = &
2\sin\theta_k\cos\varphi\,\epsilon_{xz}/\left(\epsilon_{zz} -
\sin^2\theta_k\right),\\
a_2 & = & \eta_{xx}\gamma_{yy} + \eta_{yy}\gamma_{xx} -
\eta_{xy}\gamma_{yx} -
\eta_{yx}\gamma_{xy}-\beta_{xy}\beta_{yx},\\
a_1 & = & \eta_{yy}\beta_{xx} - \eta_{xy}\beta_{yx} -
\eta_{yx}\beta_{xy},\\
a_0 & = & \eta_{xx}\eta_{yy} - \eta_{xy}\eta_{yx},
\end{eqnarray}
where the coefficients have the values:
\begin{eqnarray}
\eta_{xx} & = & \epsilon_{xx} - \sin^2\theta_k\sin^2\varphi
- \epsilon^2_{xz}/\left(
\epsilon_{zz}-\sin^2\theta_k\right),\\
\eta_{xy} & = & \epsilon_{xy} +
\sin^2\theta_k\sin\varphi\cos\varphi + 
\epsilon_{xz}\epsilon_{yz}/\left(\epsilon_{zz}-\sin^2\theta_k\right),\\
\eta_{yx} & = & -\epsilon_{xy} +
\sin^2\theta_k\sin\varphi\cos\varphi -
\epsilon_{xz}\epsilon_{yz}/\left(\epsilon_{zz}-\sin^2\theta_k\right),\\
\eta_{yy} & = & \epsilon_{yy} - \sin^2\theta_k\cos^2\varphi
+ \epsilon_{yz}^2/\left(
\epsilon_{zz}-\sin^2\theta_k\right),\\
\beta_{xx} & = &
-2\sin\theta_k\cos\varphi\epsilon_{xz}/\left(
\epsilon_{zz}-\sin^2\theta_k\right),\\
\beta_{xy} & = &
\left(\epsilon_{yz}\cos\varphi-\epsilon_{xz}\sin\varphi\right)
\sin\theta_k/\left(\epsilon_{zz}-\sin^2\theta_k\right),\\
\beta_{yx} & = & -\left(\epsilon_{xz}\sin\varphi +
\epsilon_{yz}\cos\varphi\right)
\sin\theta_k/\left(\epsilon_{zz}-\sin^2\theta_k\right),\\
\gamma_{xx} & = & - 1 - \sin^2\theta_k\cos^2\varphi/
\left(\epsilon_{zz}-\sin^2\theta_k\right),\\
\gamma_{xy} & = & \gamma_{yx}
= -\sin^2\theta_k\sin\varphi\cos\varphi/
\left(\epsilon_{zz}-\sin^2\theta_k\right),\\
\gamma_{yy} & = & - 1 - \sin^2\theta_k\sin^2\varphi/
\left(\epsilon_{zz}-\sin^2\theta_k\right),
\label{eq:gammayy}
\end{eqnarray}

The fourth order polynomial defined by
Eqs.~(\ref{eq:polynj})--(\ref{eq:gammayy}) has much
better numerical properties than the eighth-order polynomial
described by Eq.~(A4) of Paper~I.   We find that a
stable, efficient method for solving
\req{eq:polynj} can be obtained by  defining the
matrix
\begin{equation}
\mathbf{M} \equiv \left(\begin{array}{cccc}
-a_3/a_4 & -a_2/a_4 & -a_1/a_4 & -a_0/a_4\\
1 & 0 & 0 & 0\\
0 & 1 & 0 & 0\\
0 & 0 & 1 & 0
\end{array}\right),
\end{equation}
and noting that the eigenvalues of $\mathbf{M}$ are equal to
the roots of \req{eq:polynj}. We use the ZGEEV
subroutine of the  LAPACK library \citep{Anderson-ea99} to
compute the eigenvalues of $\mathbf{M}$. Of the resulting
four roots, only two correspond to physical solutions  for
the transmitted waves.  To identify the correct roots, we
write the spatial  variation of the transmitted electric
field as  $E_j\,(\bm{r}) \propto
\exp\left(\mathrm{i}\omega\,\bm{n}_j\cdot\bm{r}/c\right)$. 
The amplitude of  the electric field must decay in the
transmitted wave region, leading to the condition:
\begin{equation}
\label{eq:imagcond}
\mathfrak{Im}(n_j) \le 0.
\end{equation}
At all energies and angles for the range of magnetic fields
$B=10^{12}-10^{15}$~G,  condition~(\ref{eq:imagcond})
identifies two physical solutions to 
\req{eq:polynj}. Using these values for $n_j\,$,
we write the  polarization vectors for the transmitted wave
as
\begin{eqnarray}
\etrv_j & = & \frac{1}{\sqrt{1+|K_j^\mathrm{(t)}|^2
 + |K_{z,j}^\mathrm{(t)}|^2}}\left( \begin{array}{c}
     K_j^\mathrm{(t)} \\  1  \\  K_{z,j}^\mathrm{(t)}
     \end{array}
     \right),
\label{et}
     \\
K_j^\mathrm{(t)} & = & -\frac{\gamma_{xy}n_{z,j}^2+\beta_{xy} n_{z,j} +
\alpha_{xy}}
{\gamma_{xx} n_{z,j}^2 + \beta_{xx} n_{z,j} + \alpha_{xx}},\\
K_{z,j}^\mathrm{(t)} & = & \frac{
\epsilon_{xz} - \sin\theta_k\sin\varphi n_{z,j}
-\left(\epsilon_{xz}
 + \sin\theta_k\cos\varphi n_{z,j}\right) K_j^\mathrm{(t)}
    }{
    \epsilon_{zz} - \sin^2\theta_k}.
\end{eqnarray}

\subsection{Reflectivity calculation}

Once the quantities $n_j$ and $\etrv_j$ are known, the
reflectivity of the medium can  be calculated using the
boundary conditions for Maxwell's equations at the condensed
matter surface:
\begin{eqnarray}
\label{eq:Eboundary}
\Delta \bm{E} \times {\hat\mathbf{z}} & = & \mathbf{0},\\
\label{eq:Bboundary}
\Delta \bm{B} \times {\hat\mathbf{z}} & = & \mathbf{0},
\end{eqnarray}
where
$
\Delta \bm{E} \equiv \bm{E}^\mathrm{(i)}
 + \bm{E}^\mathrm{(r)} - \bm{E}^\mathrm{(t)}
$
and
$
\Delta \bm{B} \equiv \bm{B}^\mathrm{(i)}
 + \bm{B}^\mathrm{(r)} - \bm{B}^\mathrm{(t)}
$
are the differences between the fields above (incident and
reflected) and below (transmitted) the condensed surface.  
For the  detailed forms of the fields, see \S3.1 of
Paper~I.  Writing out the components of 
(\ref{eq:Eboundary}) and (\ref{eq:Bboundary}) for the two
orthogonal linear polarizations of the  incident wave yields a system of 
equations for the amplitudes of the reflected and
transmitted modes, analogous to Eq.~(A6) of Paper~I.   This
set of equations can be solved as two independent linear
systems with complex coefficients, such that
\begin{equation}
\mathbf{C} \cdot \left(\begin{array}{cc}
r_{11} & r_{12}\\
r_{21} & r_{22}\\
t_{11} & t_{12}\\
t_{21} & t_{22}
\end{array}\right) = \left(\begin{array}{cc}
-\cos\varphi & \cos\theta_k\sin\varphi\\
-\sin\varphi & -\cos\theta_k\cos\varphi\\
-\cos\theta_k\sin\varphi & -\cos\varphi\\
\cos\theta_k\cos\varphi & -\sin\varphi
\end{array}\right),
\end{equation}
where
\bea
\mathbf{C} &=&
\left( \begin{array}{cccc}
 \cos\varphi & 
 \cos\theta_k\sin\varphi \\
 \sin\varphi & -\cos\theta_k\cos\varphi \\
-\cos\theta_k\sin\varphi & \cos\varphi \\
 \cos\theta_k\cos\varphi
& \sin\varphi
 \end{array} \right.
\nonumber\\&&\qquad
\left. \begin{array}{cccc}
-1  & -1,\\
\etr_{1,x} & \etr_{2,x},\\
 \sin\theta_k\cos\varphi \etr_{1,z} - n_{z,1} \etr_{1,x} &
\sin\theta_k\cos\varphi \etr_{2,z} - n_{z,2} \etr_{2,x}\\
 \sin\theta_k\sin\varphi \etr_{1,z} - n_{z,1} &
 \sin\theta_k\sin\varphi \etr_{2,z} - n_{z,2}
 \end{array} \right)
\nonumber\eea 
and
$
\etr_{j,x} = \etrv_j \cdot {\hat\mathbf{x}},$
$\etr_{j,z} = \etrv_j \cdot {\hat\mathbf{z}}.
$
We solve the complex systems using the ZGESV subroutine 
of the LAPACK library \citep{Anderson-ea99}.

The corrected results for the case of a magnetized iron surface (to
be compared with Paper~I) are presented in Sect.~\ref{sect:reflGen}.

\section{Approximations for reflectivities
at the bottom of a thin atmosphere}
\label{sect:rmj}

\subsection{Reflectivities of the normal modes
in terms of $r_{mj}$}
\label{sect:RMM'}

In general, the interface between the thin atmosphere and
magnetic condensed surface has reflection and transmission
properties that are different from those of the condensed surface in
vacuum. Therefore, a separate calculation of the reflectivity
coefficients $r_{mj}$ is needed for every set of atmosphere
parameters. However, assuming that the atmosphere is
sufficiently rarefied, we may approximately replace these
coefficients by those in vacuum.
Under these conditions,
the plane waves in the atmosphere
are almost transverse, so we can approximately  set
$K_{z,M}\to0$ in \req{e'}. Then each incident and reflected
wave can be expanded over the linear polarization vectors
$\bm{e}_1$ and $\bm{e}_2$ that have been employed in the
reflectivity calculation.

For the incident (i) and reflected (r) beams, we define orthonormal
vectors
$\bm{e}_1^\mathrm{(i,r)}=\hat{\mathbf{z}}\times\bm{k}/|\hat{\mathbf{z}}\times\bm{k}|
=\hat{\mathbf{z}}\times\hat{\mathbf{k}}_\mathrm{i,r}/|\sin\theta_k|$,
$\bm{e}_2^\mathrm{(i)}=\hat{\mathbf{k}}_\mathrm{i}
  \times\bm{e}_1^\mathrm{(i)}$, and
$\bm{e}_2^\mathrm{(r)}=\bm{e}_1^\mathrm{(r)}
  \times\hat{\mathbf{k}}_\mathrm{r}$, where 
$\hat{\mathbf{k}}$
denotes the unit vector along $\bm{k}$.
In the notations of Fig.~\ref{fig:plot1},
\bea&&\!\!
   \hat{\mathbf{k}}_\mathrm{i,r} =
       \sin\theta_k\cos\varphi \,\hat{\mathbf{x}}
       + \sin\theta_k\sin\varphi \, \hat{\mathbf{y}}
       \mp\cos\theta_k \, \hat{\mathbf{z}},
\\&&\!\!
   \bm{e}_1^\mathrm{(i)} = \bm{e}_1^\mathrm{(r)}
    =  -\sin\varphi \,\hat{\mathbf{x}}
    + \cos\varphi  \,\hat{\mathbf{y}},
\\&&\!\! 
   \bm{e}_2^\mathrm{(i,r)} = 
       \cos\theta_k \,
       (\cos\varphi \,\hat{\mathbf{x}}
       + \sin\varphi  \,\hat{\mathbf{y}} )
       \pm \sin\theta_k\, \hat{\mathbf{z}},
\eea
where the upper and lower signs in $\mp\cos\theta_k$ are for the
incident and reflected waves, respectively.
The coordinates in which
\req{e'} is written are $(x',y',z')$ (Fig.~\ref{fig:plot1}),
defined according to relations
$\hat{\mathbf{y'}}=\bm{B}\times\bm{k}/|\bm{B}\times\bm{k}|$
and 
$\hat{\mathbf{x'}}=\hat{\mathbf{y'}}\times\hat{\mathbf{k}}$. 

The electric field
of the incoming ray with unit amplitude and polarization
$M'$ ($M'=$X or $M'=$O) can be written as
\beq
  \bm{e}^\mathrm{(i)}_{M'} = 
     c_{M'1}^\mathrm{(i)} \bm{e}_1^\mathrm{(i)}
      + c_{M'2}^\mathrm{(i)} \bm{e}_2^\mathrm{(i)},
\label{cM'}
\eeq
where $c_{M'j}^\mathrm{(i)} = \bm{e}_{M'}^\mathrm{(i)}
\cdot \bm{e}_j^\mathrm{(i)}$.
According to Eqs.~(\ref{def_r}) and (\ref{cM'}), the
reflected field is 
\beq
   r'_{\mathrm{X}M'} \bm{e_\mathrm{X}^\mathrm{(r)}}
   + r'_{\mathrm{O}M'} \bm{e_\mathrm{O}^\mathrm{(r)}}
   =
   \sum_{m=1}^{2} \sum_{j=1}^{2} r_{mj} 
      c_{M'j}^\mathrm{(i)} \bm{e}_m^\mathrm{(r)} .
\eeq
The amplitudes $r'_{\mathrm{X}M'}$ and
 $r'_{\mathrm{O}M'}$ of the reflected-field components in
the X- and O-modes, respectively, are given by the solution
of the linear system
\beq
   \left( \begin{array}{cc}
      c_\mathrm{X1}^\mathrm{(r)}  & c_\mathrm{O1}^\mathrm{(r)}
 \\
\rule{0pt}{3ex}
      c_\mathrm{X2}^\mathrm{(r)}  & c_\mathrm{O2}^\mathrm{(r)}
    \end{array} \right)
   \left( \begin{array}{cc}
        r'_{\mathrm{XX}} & r'_{\mathrm{XO}} 
 \\
\rule{0pt}{2.5ex}
 r'_{\mathrm{OX}} & r'_{\mathrm{OO}}
    \end{array} \right)
       =
   \left( \begin{array}{cc}
      r_{11} & r_{12} \\
      r_{21} & r_{22}
    \end{array} \right)
   \left( \begin{array}{cc}
      c_\mathrm{X1}^\mathrm{(i)}  & c_\mathrm{O1}^\mathrm{(i)}
 \\
\rule{0pt}{3ex}
      c_\mathrm{X2}^\mathrm{(i)}  & c_\mathrm{O2}^\mathrm{(i)}
    \end{array} \right)
     ,
\label{r'MM'}
\eeq
where $c_{Mj}^\mathrm{(r)} = \bm{e}_{M}^\mathrm{(r)}
\cdot \bm{e}_j^\mathrm{(r)}$. Since
the incident X- and O-modes are incoherent,
the normal mode
reflectivities in Eqs.~(\ref{XX}) and (\ref{OO}) are
\beq
   R_{MM'} = | r'_{MM'} |^2.
\eeq
According to \req{e'},
\beq
   c_{Mj}^\mathrm{(i,r)} =
      \frac{\mathrm{i}K_M^\mathrm{(i,r)}
        \,\hat{\mathbf{x'}}_\mathrm{i,r}\cdot\bm{e}_j^\mathrm{(i,r)}
         + \hat{\mathbf{y'}}_\mathrm{i,r}\cdot\bm{e}_j^\mathrm{(i,r)}
         }{
         \sqrt{1+|K_M^\mathrm{(i,r)}|^2}} ,
\label{cMj}
\eeq
where $K_M^\mathrm{(i,r)}=K_M(\alpha_\mathrm{i,r})$.
The explicit expressions for the scalar products in
\req{cMj} are\footnote{We thank Denis Gonzalez Caniulef for finding an
error in the first version of Eq.(\ref{y1e2}).}
\bea
   \hat{\mathbf{x_{\,\mathrm{i,r}}'}}\cdot\bm{e}_1^\mathrm{(i,r)}
     &=&
      \sin\theta_B\sin\varphi/\!\sin\alpha_\mathrm{i,r},
\\
   \hat{\mathbf{y_{\mathrm{i,r}}'}}\cdot\bm{e}_1^\mathrm{(i,r)} &=&
      (\cos\theta_B\sin\theta_k \pm
      \sin\theta_B\cos\theta_k\cos\varphi)/\!\sin\alpha_\mathrm{i,r},
\\
   \hat{\mathbf{x_{\,\mathrm{i,r}}'}}\cdot\bm{e}_2^\mathrm{(i,r)} &=&
      (\mp\cos\theta_B\sin\theta_k -
      \sin\theta_B\cos\theta_k\cos\varphi) /
        \!\sin\alpha_\mathrm{i,r},
\\
   \hat{\mathbf{y_{\mathrm{i,r}}'}}\cdot\bm{e}_2^\mathrm{(i,r)} &=&
     \pm \sin\theta_B\sin\varphi / \sin\alpha_\mathrm{i,r}.
\label{y1e2}
\eea

Caution should be used when employing approximations for
$r_{mj}$ in \req{r'MM'} if one of the normal
modes is almost completely reflected, that is, $R_{MM'}\approx1$.
Such a situation occurs for the X-mode at $\Eci<E<\EC$, if
both $\bm{k}$ and $\bm{B}$ are close to normal (see
Paper~I). In this case the fit error may exceed
$(1-R_\mathrm{XX})$ and result in $R_\mathrm{XX}>1$, which
is unphysical. In particular, the fitting formulae presented
below may occasionally give $R_\mathrm{XX}$ a few percent
above 1 at very small $\theta_B$ and $\theta_k$. In such
instances one should truncate the mode-specific
reflectivities, recovered from the fit, so as to fulfill the
general condition $0<R_{MM'}<1$. 

\subsection{Approximations for $r_{mj}$}

For calculating $R_{MM'}$ according to
Sect.~\ref{sect:RMM'}, we use an analytic model of the
complex reflectivity coefficients $r_{mj}$, which
agrees with the approximations derived in
Sect.~\ref{sect:reflFe} and roughly reproduces the
computed dependences of
 $r_{mj}=|r_{mj}|\exp(\mathrm{i}\phi_{mj})$
  on $E$ for many characteristic geometry settings. For the
squared moduli we use the following expressions:
\bea
   |r_{12}|^2 &=& \left\{ \begin{array}{l}
   0 \quad\textrm{in Region~I},
    \\ 
    f_E\,(1-J_\mathrm{B1})\,(1-J_\mathrm{C})   
    \\ \qquad
     + J_\mathrm{C} R_L
     (1-\sin\theta_B\sin^2\varphi)/2
       \quad\textrm{in Region~II},
    \end{array} \right.
\\
   |r_{21}|^2 &=& \left\{ \begin{array}{l}
   0 \quad\textrm{in Region~I},
    \\ 
    f_E\,(1+J_\mathrm{B1}-2J_\mathrm{B})\,(1-J_\mathrm{C})
    \\ \qquad 
       + J_\mathrm{C} R_L \sin\theta_k\,(1-\cos\alpha)/2
    \quad\textrm{in Region~II},
    \end{array} \right.
\\
      |r_{11}|^2 &=& 1-J_1-|r_{12}|^2,
\quad
      |r_{22}|^2 = 1-J_2-|r_{21}|^2,
\eea
and $ f_E \equiv E/(E+\tEC/2) $.
The functions
$J_\mathrm{1}$, $J_\mathrm{2}$, $J_\mathrm{B}$,
$J_\mathrm{C}$, 
$J_\mathrm{B1}$,
and $R_L$ are
defined in Sect.~\ref{sect:reflFe}.
In the case of the free-ions model,
small accidental discontinuities
at the boundary of Region~I are eliminated by truncating
$|r_{11}|^2$ and $|r_{22}|^2$ from above
by their values at $E=\Eci$.
Our approximations for the complex phases are
\bea
   \phi_{11} &=& \left\{ \begin{array}{l}
       \pi \quad\textrm{in Region~I},
       \\ 
       -\pi f_L,
       \quad\textrm{if}\quad E>\tEC,
       \\
       \displaystyle{
       \pi + \pi\,\frac{E-\Eci}{\tEC-\Eci}
       }
       \quad\textrm{otherwise},
    \end{array} \right.
    \\
    \phi_{22} &=& \left\{ \begin{array}{l}
       \pi f_L,
       \quad\textrm{if}\quad E>\tEC,
       \\
       \phi_{11}
       \quad\textrm{otherwise},
    \end{array} \right.
    \\
    \phi_{12} &=&  \left\{ \begin{array}{l}
       -\pi/2 \quad\textrm{in Region~I},
       \\ 
       \pi/2 - 2\pi f_L,
       \quad\textrm{if}\quad E>\tEC,
       \\
       {\displaystyle
       \pi\,\frac{E-(\Eci+\tEC)/2}{\Eci-\tEC}
       }
       \quad\textrm{otherwise},
    \end{array} \right.
    \\
    \phi_{21} &=& \phi_{12}+\pi,
\eea
where
\[
    f_L =
    \left[1+\exp\left( 5\,
         \frac{E_L-E}{E_L-\tEC}
          \right)\right]^{-1}.
\]

\end{document}